\documentclass[final,3p,12pt,times,preprint]{elsarticle}
\usepackage{amssymb}
\usepackage{atlasphysics}
\usepackage{float}
\newcommand\T{\rule{0pt}{2.6ex}}
\newcommand\B{\rule[-1.2ex]{0pt}{0pt}}
\journal{Nuclear Physics B}

\begin{document}

\begin{frontmatter}
\hfill{\footnotesize{CERN-OPEN-2014-039;\ \ FTUAM-14-28}}
\vspace{-0.2cm}
\title{Determination of the $b$-quark mass $m_b$ from the angular screening effects in the ATLAS $b$-jet shape data.}
\tnotetext[label1]{}
\author[]{Javier Llorente} 
\ead{javier.llorente.merino@cern.ch}

\author[]{Josu Cantero}
\ead{josu.cantero.garcia@cern.ch}

\address{Universidad Aut\'onoma de Madrid (UAM), Facultad de Ciencias.\\
Departamento de F\'isica Te\'orica. Cantoblanco, Madrid 28049, Spain}

\begin{abstract}
The dependence of jet shapes in $t\bar{t}$ events on the $b$-quark mass and the strong coupling is investigated. To this end, the \textsc{Pythia} Monte Carlo generator is used to produce samples of $t\bar{t}$ events in $pp$ collisions at $\sqrt{s} = 7 \TeV$, performing a scan over the values for the shower QCD scale $\Lambda_s$ and the $b$-quark mass $m_b$. The obtained jet shapes are compared with recently published data from the ATLAS Collaboration. From fits to the light-jet data, the Monte Carlo shower scale is determined, while the $b$-quark mass is extracted using the $b$-jet shapes. The result for the mass of the $b$-quark is $m_b = 4.86 ^{+0.49}_{-0.42}\GeV$.
\end{abstract}


\end{frontmatter}

\section{Introduction}
\label{secIntro}
It is a well established fact that hadrons produced in $e^+e^-$, $ep$ and $pp$ colliders at high momentum transfers appear in well collimated bundles called jets. These jets are understood to proceed via a two step process. The first one, which is of a perturbative nature, relates to the formation of a parton shower following the underlying hard partonic interaction. The second, which is non-perturbative, is called hadronisation and relates to the way partons in the shower recombine to form colourless hadrons. Hadronisation effects are expected to become smaller at higher transverse momentum scales.\\
\newline
Jet shapes \cite{ellis,vitev} are defined as the normalised transverse momentum flow as a function of the distance to the jet axis. They are considered to be a measure of the jet internal structure. Thus, at high energies, they are sensitive to the amount of final state radiation.\\
\newline
Recently, the ATLAS Collaboration has published data on $b$-jet and light jet shapes measured in $t\bar{t}$ final states \cite{atlasData}. Here $b$-jets arise from the decays $t\rightarrow W b$ in both the single-lepton and dilepton modes. Light jets are studied in the single lepton channel, where one $\Wboson$ is decaying leptonically and the second one hadronically. It is found that $b$-jets are broader than light jets.\\
This is understood to be due to the fact that the angular radiation pattern for a $b$-quark is significantly different than that of a light-quark due to the heavier mass of the former. These effects were thoroughly studied in Ref. \cite{marchesini} for the full angular range subtended from the direction of motion of the $b$-quark.\\
\newline
Indeed, for a parton branching $q \rightarrow \tilde{q}g$, the invariant mass of the decay products can be written as $m_q^2 \simeq 2E_{\tilde q}E_g(1-\cos\theta)$ in the regime where the quark mass is negligible compared to its energy scale. Here $\theta$ is the angle formed by the 3-momenta of the final-state quark and the radiated gluon. In the collinear limit, valid for the jet cone region, one can expand the cosine as a Taylor series and easily obtain the relation
\newline
\begin{equation}
\theta \simeq \frac{m_q}{\sqrt{E_{\tilde q} E_g}} = \frac{1}{\sqrt{z(1-z)}}\frac{m_q}{E_q}
\label{eq:angleMass}
\end{equation}
\newline
Here, $z$ is the fraction of energy carried by the gluon ($E_g = z E_q$). Eq. \ref{eq:angleMass} suggests that there is a direct relationship between the mass of the branching parton and the angular distribution of the resulting products around the jet axis. For light-quark jets, the dominant effect on the opening angle described by Eq. \ref{eq:angleMass} arises from the gluon energy fraction $0 < z < 1$. On the other hand, the opening angle in $b$-jets is controlled by the heavier mass of the $b$-quark.\\
\newline
Defining $\theta_0 = m_q/E_q$, the probability of a gluon emission at a small opening angle $\theta < \theta_0 \muchless 1$ is given by \cite{dokshitzer}
\newline
\begin{equation}
\left(\frac{d\sigma}{d\omega}\right)_{q\rightarrow\tilde{q}g} = \frac{\alpha_s C_F}{\pi\omega}\frac{(2\sin\theta/2)^2 d(2\sin\theta/2)^2}{[(2\sin\theta/2)^2+\theta_0^2]^2}\left[1+\mathcal{O}(\theta_0,\omega)\right]
\sim \frac{1}{\omega}\frac{\theta^2d\theta^2}{[\theta^2+\theta_0^2]^2}
\label{eq:gluonProb}
\end{equation}
\newline
In Eq. \ref{eq:gluonProb}, $\omega$ corresponds to the energy of the radiated gluon. From here one can infer that for the kinematical region with $\theta < \theta_0$ the amount of radiation is highly suppressed. This effect is known as angular screening, and the region $\theta < \theta_0$ is known as the `dead cone'.\\
\newline
This discussion proves interesting to investigate the dependence of the $b$-jet shapes on the $b$-quark mass. This is the purpose of this paper. To this end, the \textsc{Pythia} Monte Carlo program \cite{pythia} was used to generate samples of $t\bar{t}$ events where both the shower scale $\Lambda_s$ and the $b$-quark mass $m_b$ were varied in the ranges $[20 , 300] \MeV$ and $[4 , 6] \GeV$ respectively. In this study, only the first three $\pt$ bins studied in \cite{atlasData} are introduced into the fits. This is done to maximise the effect of the $b$-quark mass in the jet shapes, which is largely reduced at high $\pt$ because of the inverse proportionality of $\theta_0$ with the energy of the parent quark. The outline of the paper is as follows: the MC predictions are discussed in Sect. \ref{secPythia}. In Sect. \ref{secJetSel} the jet selection and the jet shape definition are discussed. The fitting procedure is addressed in Sect. \ref{secFit}, while Sections \ref{secLambda} and \ref{secMb} are dedicated to the extraction of $\Lambda_s$ and $m_b$, respectively. In Section \ref{secUnc}, the theoretical uncertainties are described. Finally, Section \ref{secResults} is left for summary and conclusions.\\
\newpage
\section{Monte Carlo predictions}
\label{secPythia}
Top-quark pair events have been generated using the \textsc{Pythia 6.4} program. Additionally, the \textsc{MSTJ(42)=3} switch has been used to take into account the larger mass of the $b$-quark on the angular distribution of the decay products \cite{norr01}. Also, the switch \textsc{MSTJ(43)=3} has been used to set the fragmentation variable $z$ as the fraction of energy in the centre-of-mass frame of the showering partons \cite{pythia}.\\
\newline
Jet shapes naturally depend on the strong coupling constant $\alpha_s$, as it controls the radiation emitted by strongly-interacting partons, and have been in fact a precise way to determine its value in Ref. \cite{zeus}. Therefore, one needs to take this effect into account for a precise determination of the $b$-quark mass. At the one-loop order, the scale dependence of the strong coupling can be parametrised by \cite{pdg}
\newline
\begin{equation}
\alpha_s(Q^2) = \frac{1}{\beta_0\log\left(\frac{Q^2}{\Lambda^2}\right)}; \ \ \beta_0 = \frac{1}{4\pi}\left(11-\frac{2}{3}n_f\right)
\label{eq:asLO}
\end{equation}
\newline
Eq. \ref{eq:asLO} incorporates the QCD scale $\Lambda$, which can be varied for the \textsc{Pythia} time-like parton showers arising from a resonant decay using the \textsc{PARJ(81)} switch. Finally, the $b$-quark mass $m_b$ is varied around its nominal value $m_b = 4.8 \GeV$ using the \textsc{PMAS(5)} and \textsc{PARF(105)} switches, which control the kinematical mass of the $b$-quark and its constituent mass, respectively. Additionally, $t\bar{t}$ samples have been generated using the \textsc{Herwig++} Monte Carlo program \cite{herwigpp}. The differences between the value of $m_b$ obtained in \textsc{Herwig++} and that obtained using \textsc{Pythia} will be discussed later, and assigned as a theoretical uncertainty.
\section{Jet selection and jet shape calculation}
\label{secJetSel}
The final-state particles from the \textsc{Pythia} simulation are clustered using the anti-$k_t$ algorithm \cite{antiKt} as implemented in \textsc{FastJet} \cite{fastjet}, with a radius parameter $R = 0.4$. As specified in Ref. \cite{atlasData}, muons and neutrinos are left out of the clustering algorithm.\\
\newline
All jets with transverse momentum $\pt > 30 \GeV$ are pre-selected. To select the jets induced by $b$-quarks from the top decays, a matching procedure is used between the clustered jets and any hadron containing $b$-quarks. If one of these hadrons with $\pt > 5 \GeV$ is found at a distance $\Delta R = \sqrt{(\Delta\eta)^2 + (\Delta\phi)^2} < 0.3$ from the axis of a given jet, this jet is selected as a $b$-jet. Alternatively, light-quark jets are selected as the pair of jets which, not containing $B$-hadrons closer than $\Delta R = 0.3$ to the jet axis, have the closest invariant mass to the nominal $\Wboson$ boson mass $m_\Wboson = 80.4 \GeV$.\\
\newline
The differential jet shape is then calculated for both samples following the formula in \cite{atlasData}
\newline
\begin{equation}
\langle\rho(r)\rangle = \frac{1}{\Delta r}\frac{1}{N_{\mathrm{jets}}}\sum_{\mathrm{jets}}\frac{p_{\mathrm{T}}(r-\Delta r/2,r+\Delta r/2)}{p_{\mathrm{T}}(0,R)}
\end{equation}

\section{Analysis procedure}
\label{secFit}
As $b$-jet shapes depend on both the parton shower QCD scale $\Lambda_s$ and the $b$-quark mass $m_b$, both need to be determined for a precise result. A simultaneous determination of both parameters is not possible because a variation of one of them can be compensated by an opposite variation of the other one, leading to a set of degenerate minima in the plane $(m_b, \Lambda_s)$. However, it is expected that the light-jet shapes in \cite{atlasData} depend only in $\Lambda_s$ and not in $m_b$. Therefore, one can determine the parameter $\Lambda_s$ from the light-jet shapes and use it for the extraction of $m_b$ from the $b$-jet data.\\
\newline
The method used for the extraction of a physical parameter $\beta = \Lambda_s, m_b$ from a theoretical distribution scan relies on the minimisation of a standard $\chi^2$ for each $\pt$ bin using \textsc{Minuit} \cite{minuit}. The $\chi^2$ function is defined in a way which takes into account the correlations between the experimental uncertainties via a set of nuisance parameters $\{\lambda_i\}$. In terms of the parameter $\beta$ to be extracted and the nuisance parameter vector $\vec{\lambda}$, it can be written as
\newline
\begin{equation}
\chi^2(\beta; \vec{\lambda}) = \sum_{k} \frac{(x_k - F_k(\beta; \vec{\lambda}))^2}{\Delta x_k^2 + \Delta \tau_k^2} + \sum_{i} \lambda_i^2
\label{eq:chi2A}
\end{equation}
\begin{equation}
F_k(\beta; \vec{\lambda}) = \phi_k(\beta)\left(1+\sum_{i}\lambda_i \sigma_{ik}\right)
\label{eq:chi2B}
\end{equation}
\newline
In Eq. \ref{eq:chi2A}, the index $k$ runs over all $r$ bins in a given $\pt$ bin, with a given value $x_k$ of the jet shape and with statistical uncertainty $\Delta x_k$. Here, $\Delta\tau_k$ represents the statistical uncertainty on the theoretical predictions. The nuisance parameters $\lambda_i$, one for each source of uncertainty, are also involved in Eq. \ref{eq:chi2B}, where the functions $\phi_k(\beta)$ correspond to the nominal dependence of the jet shape with the parameter $\beta$ in bin $k$. They are parametrised in terms of a parabola throughout this paper. Finally, $\sigma_{ik}$ are the relative uncertainties for source $i$ in the bin $k$ \cite{hepData}.\\
\newline
Each nuisance parameter corresponds to a different uncertainty on the data. Table \ref{tab:nuisPar} shows the identification of each $\lambda_i$ with the corresponding source, ordered from larger to smaller impact.

\begin{table}[H]
\caption{Identification of the nuisance parameters $\lambda_i$ with the sources of experimental uncertainty in the ATLAS data.}
\label{tab:nuisPar}
\begin{center}
\begin{tabular}{cccc}
\hline
\T \B Nuisance parameter & Source of uncertainty & Impact on data\\
\hline
\T $\lambda_1$ & Pileup & $2\%-10\%$\\
\T $\lambda_2$ & Cluster systematics & $2\%-10\%$\\
\T $\lambda_3$ & Unfolding-modelling & $1\%-8\%$\\
\T $\lambda_4$ & Jet energy scale & $\simeq 5\%$\\
\T $\lambda_5$ & Jet energy resolution & $\simeq 5\%$\\
\T\B $\lambda_6$ & JVF & $< 1\%$\\
\hline
\end{tabular}
\end{center}
\end{table}

\section{Determination of the parton shower scale $\Lambda_s$}
\label{secLambda}
\begin{figure}[H]
\centering
\includegraphics[width=16.0cm,height=14.0cm]{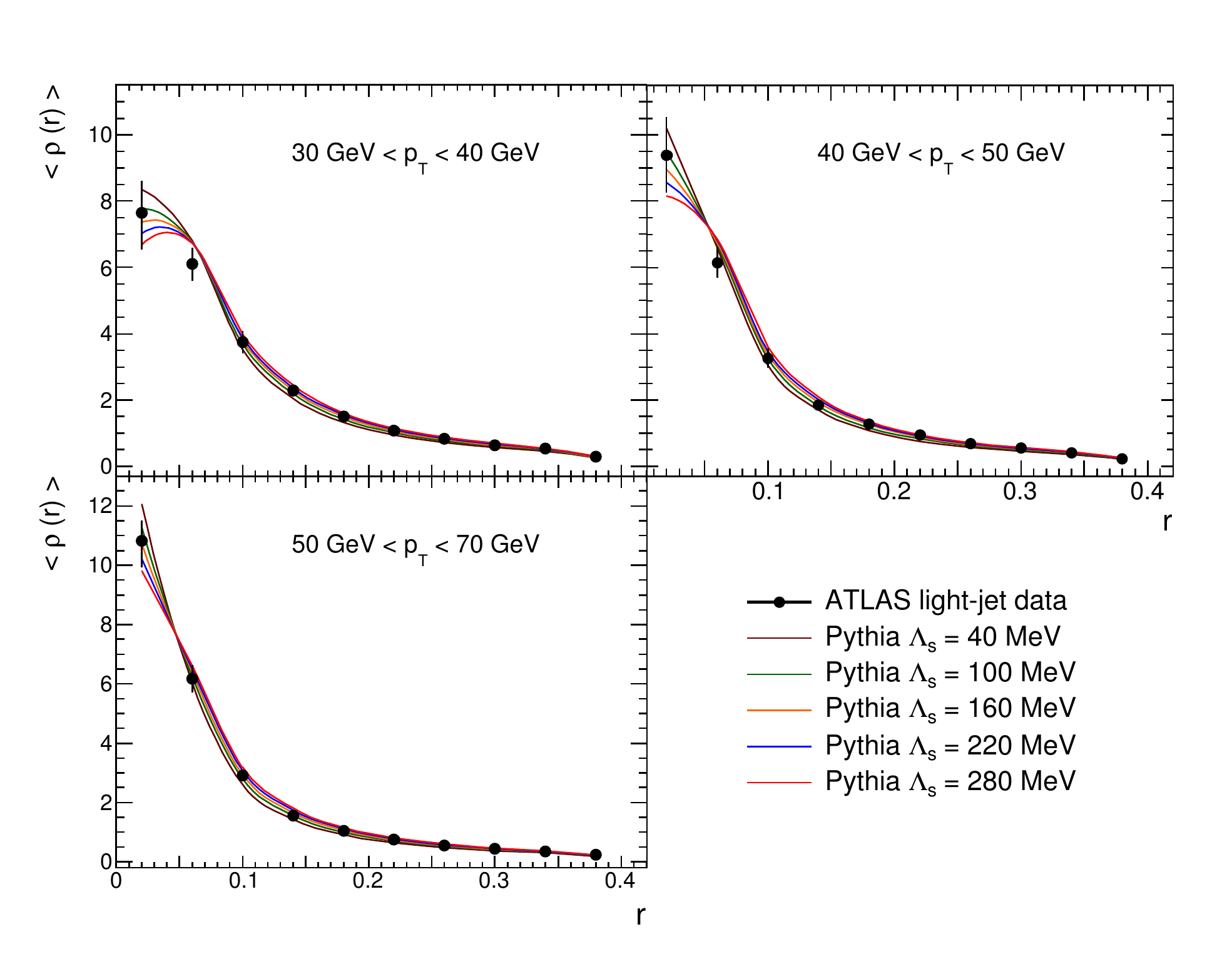}
\caption{Results of the $\Lambda_s$ scan compared to the ATLAS light-jet data in \cite{atlasData}}
\label{fig:lambdaScan}
\end{figure}

\begin{figure}[H]
\centering
\includegraphics[width=5.5cm,height=4.5cm]{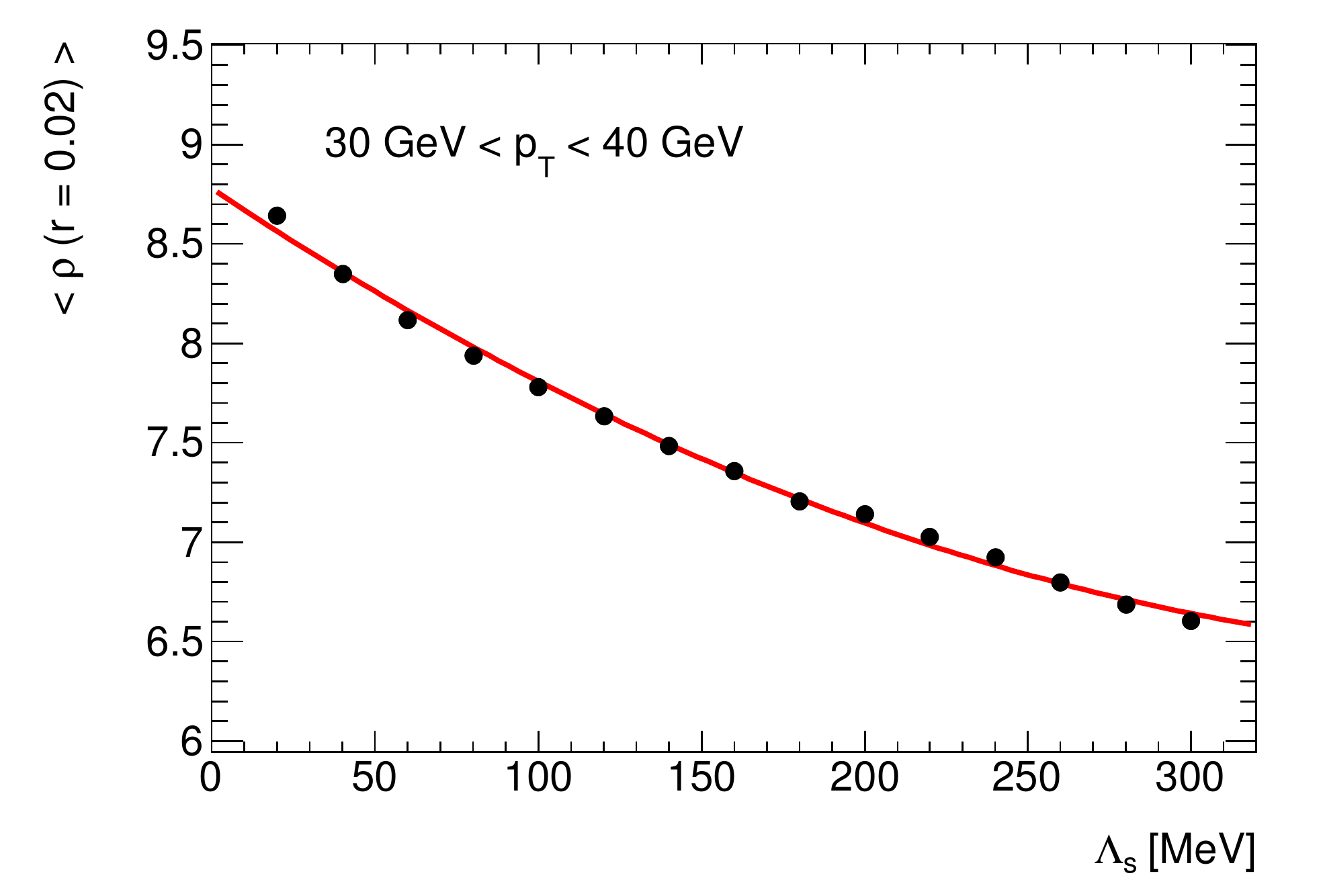}
\hspace{-0.25cm}
\includegraphics[width=5.5cm,height=4.5cm]{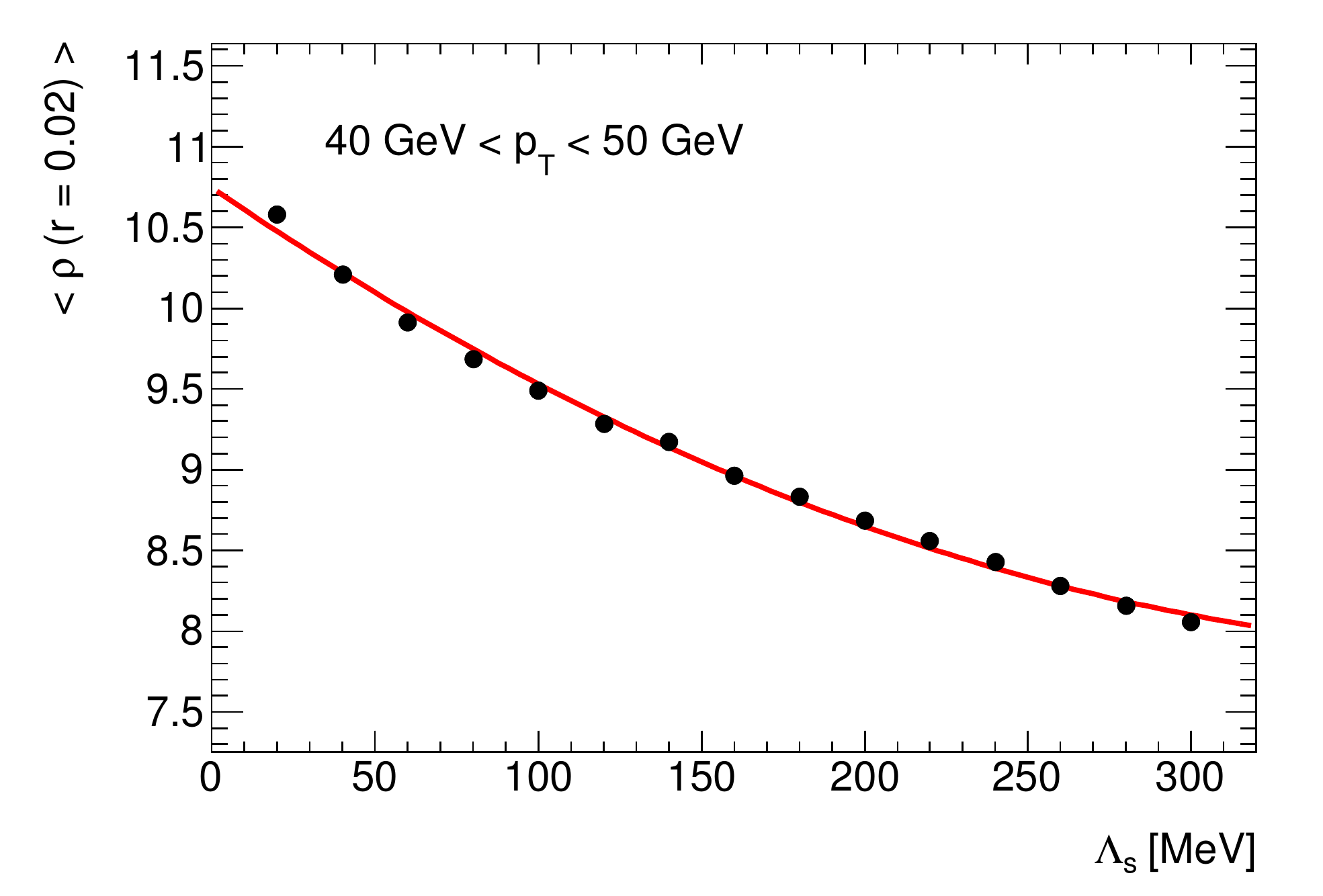}
\hspace{-0.25cm}
\includegraphics[width=5.5cm,height=4.5cm]{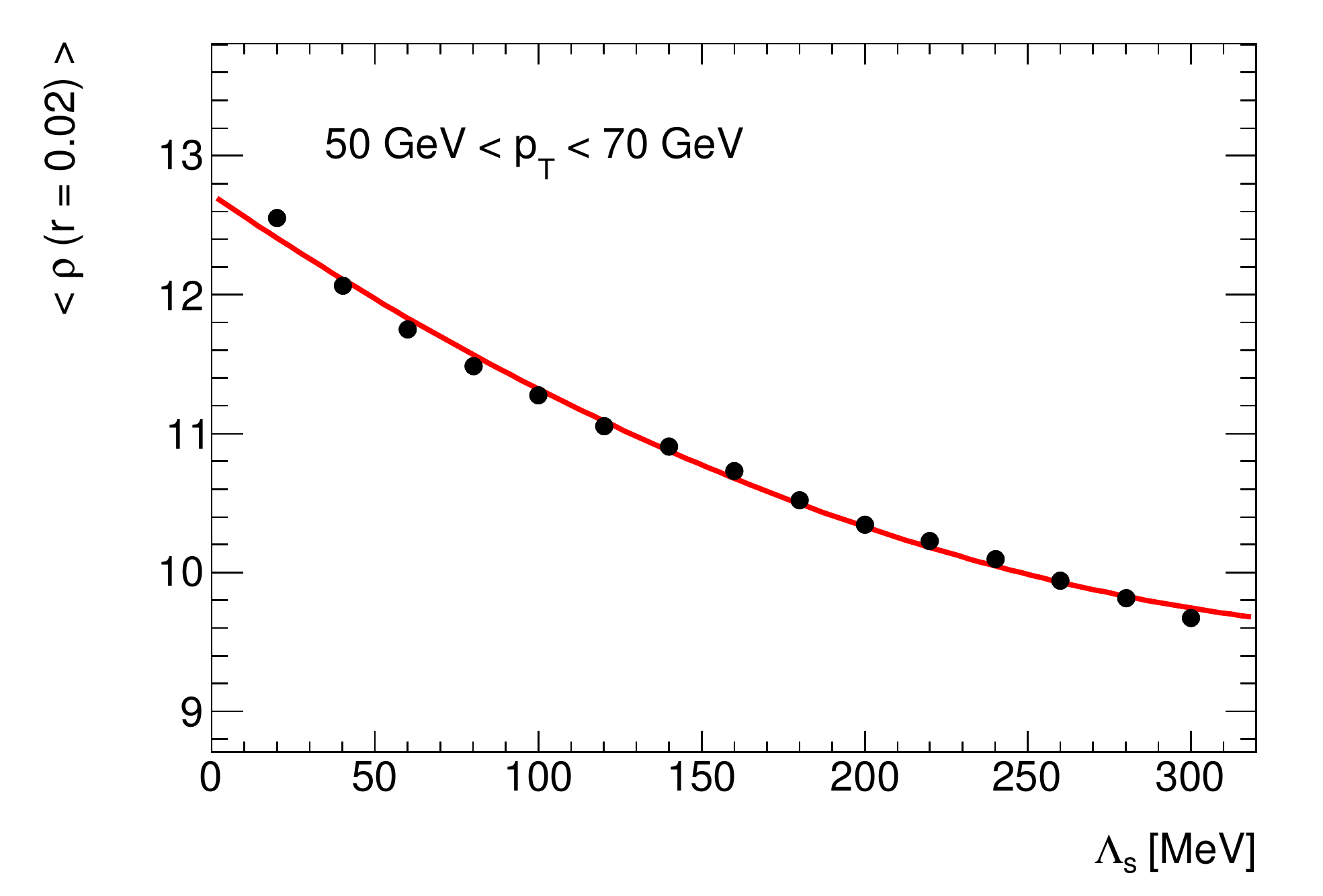}
\vspace{-0.7cm}
\caption{Dependence of the light-quark jet shape $\langle\rho(r = 0.02)\rangle$ with the parton shower scale $\Lambda_s$ for the $\pt$ intervals $30 \GeV < \pt < 40 \GeV$ (left), $40 \GeV < \pt < 50 \GeV$ (middle) and $50 \GeV < \pt < 70 \GeV$ (right), together with the interpolating functions $\phi_k(\Lambda_s)$.}
\label{fig:lambda1}
\end{figure}

\begin{figure}[H]
\centering
\includegraphics[width=6.7cm,height=5.0cm]{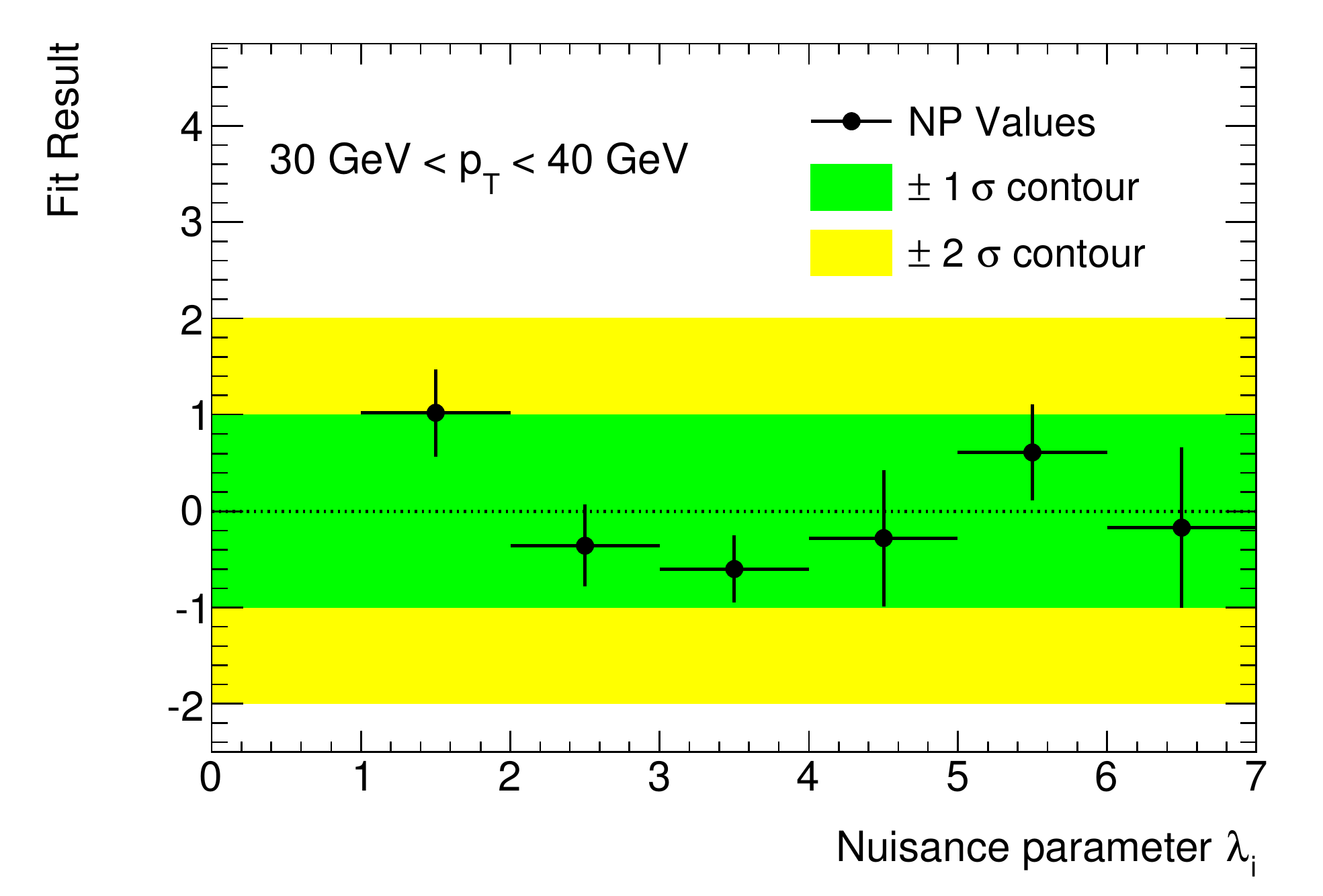}
\includegraphics[width=6.7cm,height=5.0cm]{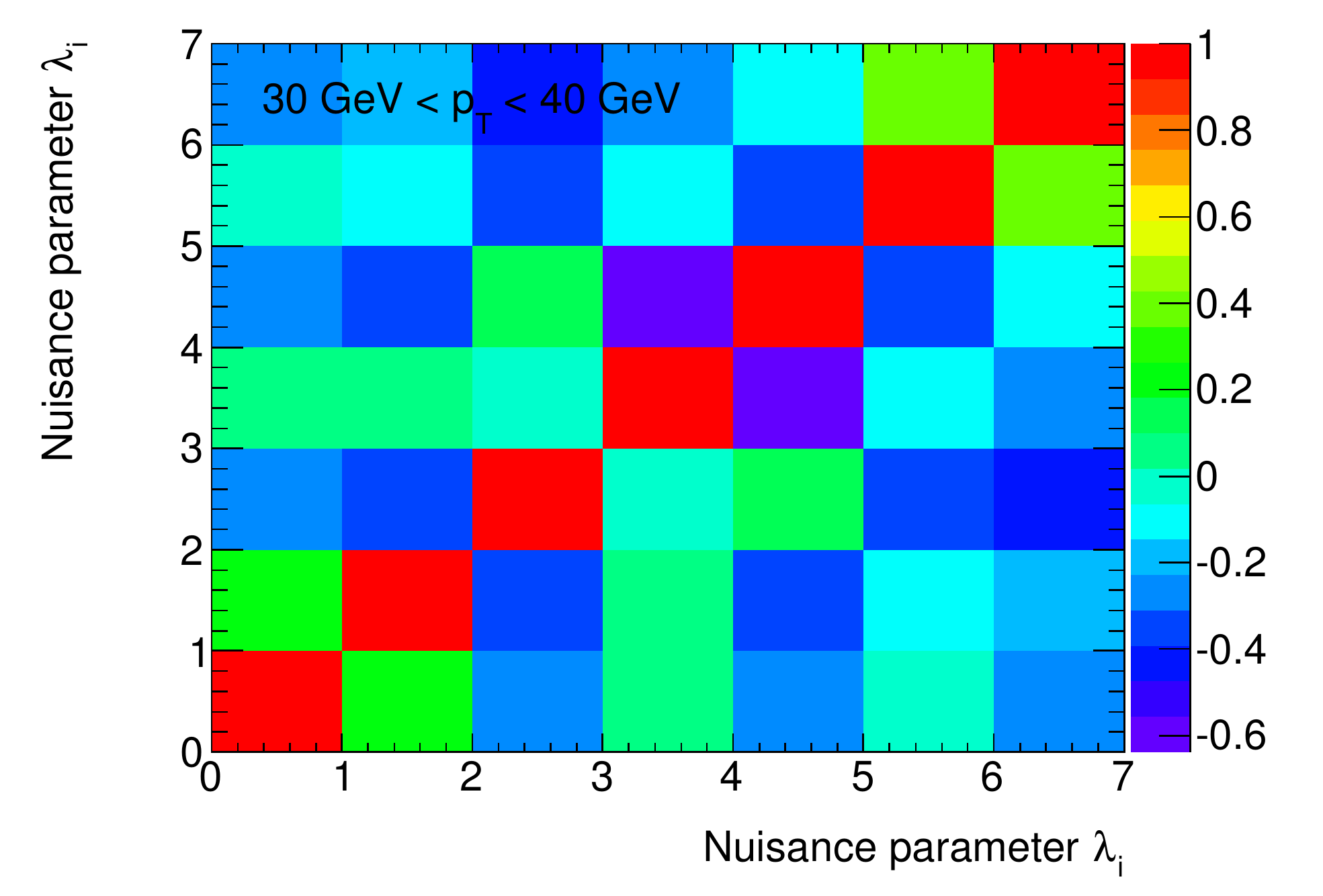}

\includegraphics[width=6.7cm,height=5.0cm]{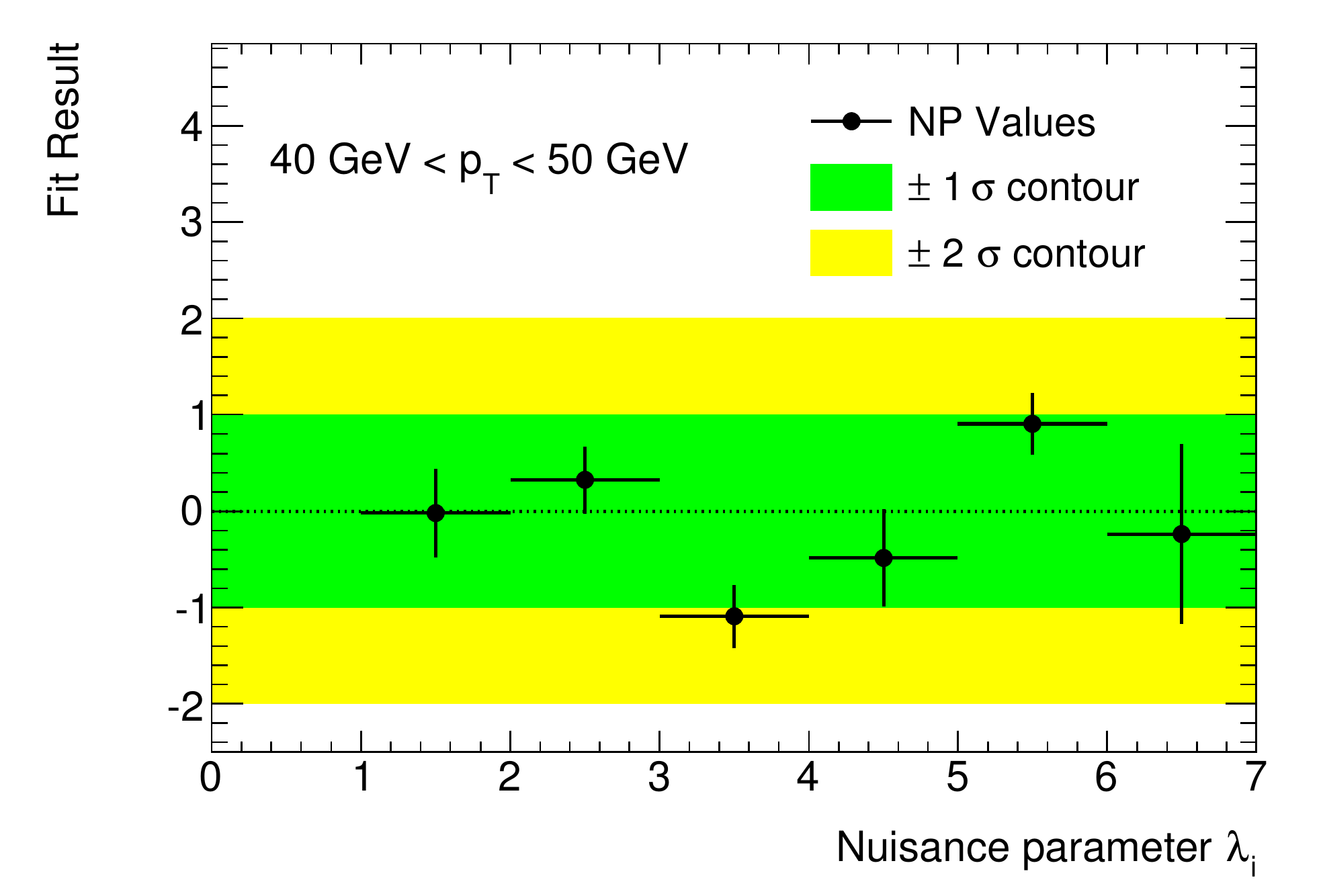}
\includegraphics[width=6.7cm,height=5.0cm]{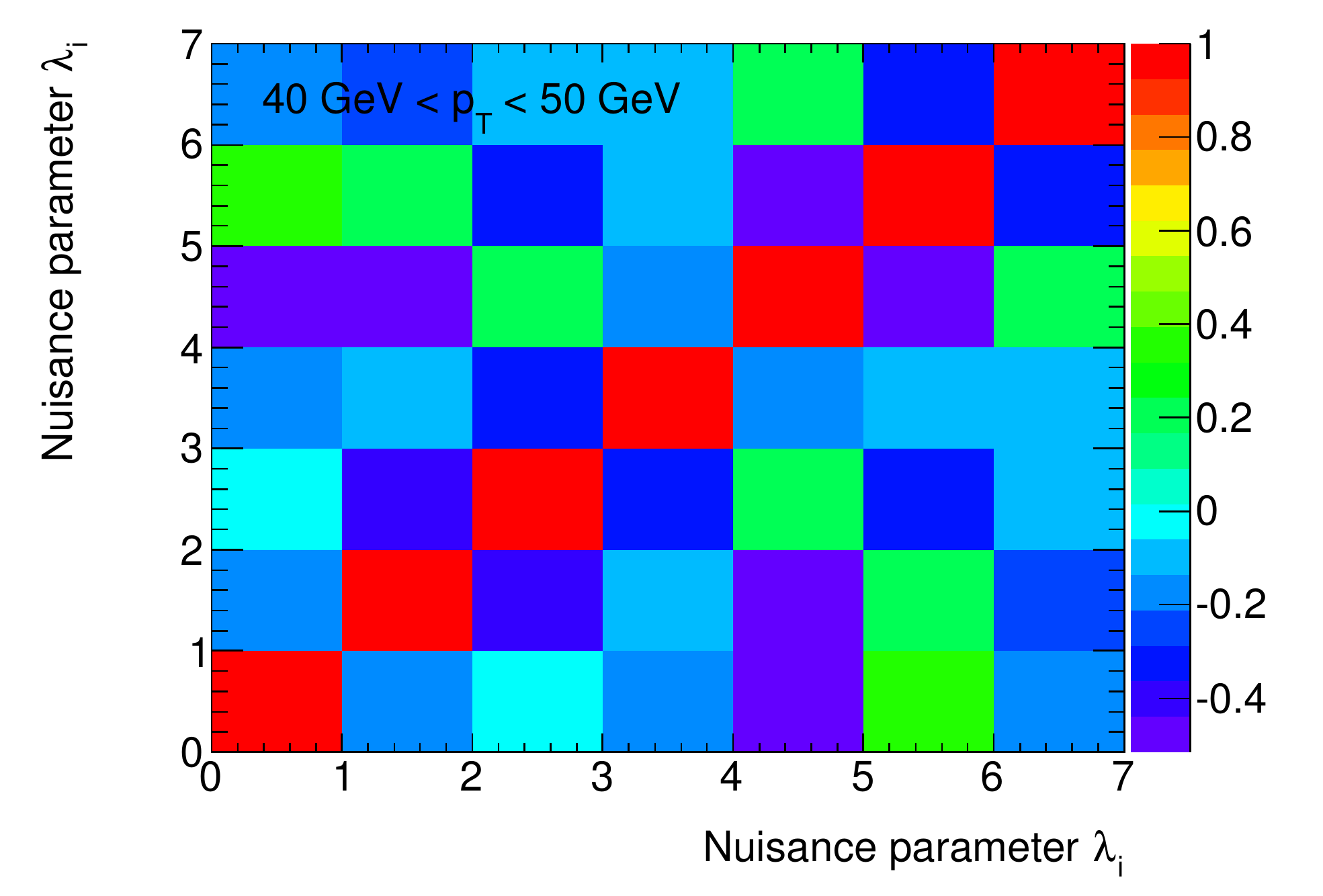}

\includegraphics[width=6.7cm,height=5.0cm]{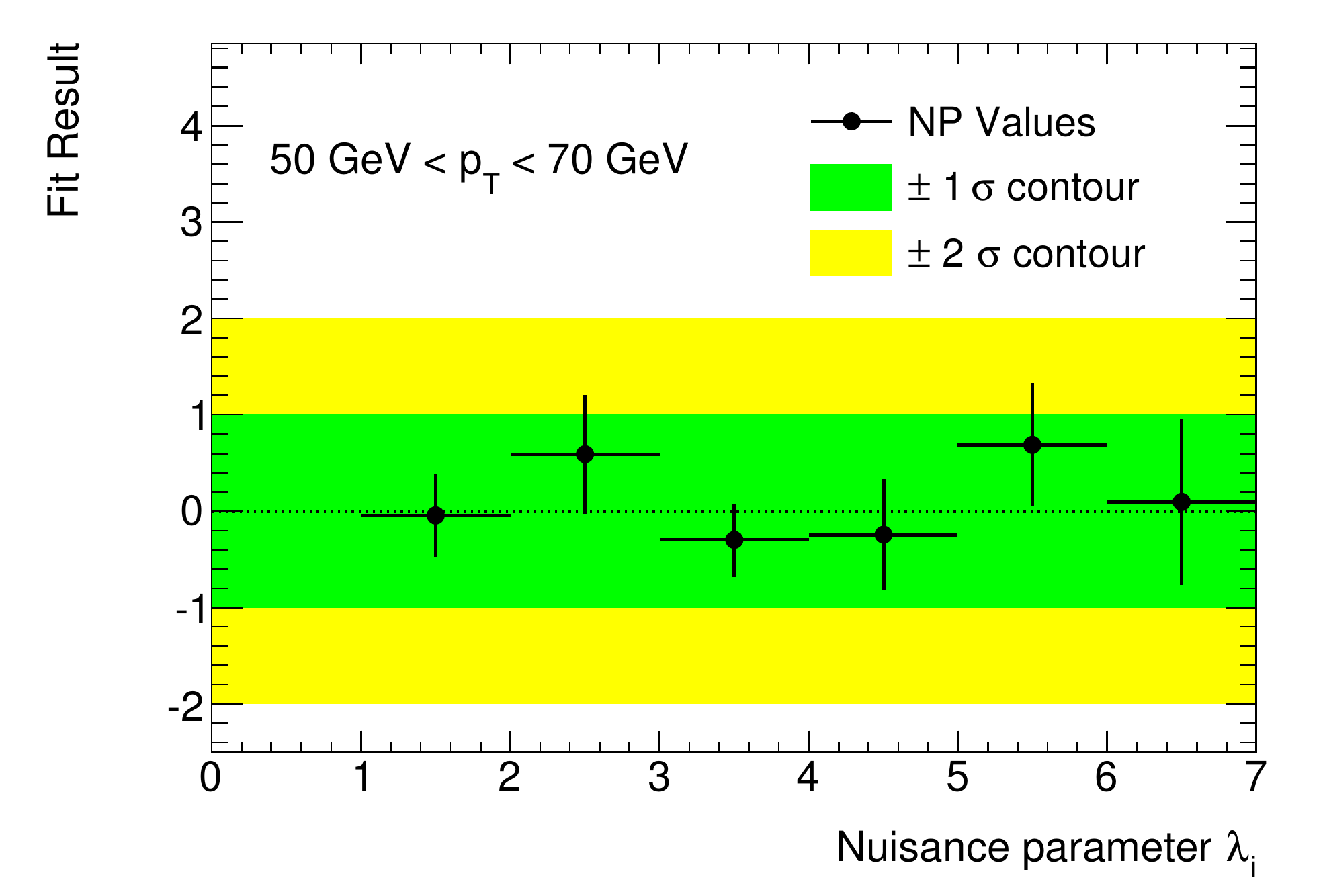}
\includegraphics[width=6.7cm,height=5.0cm]{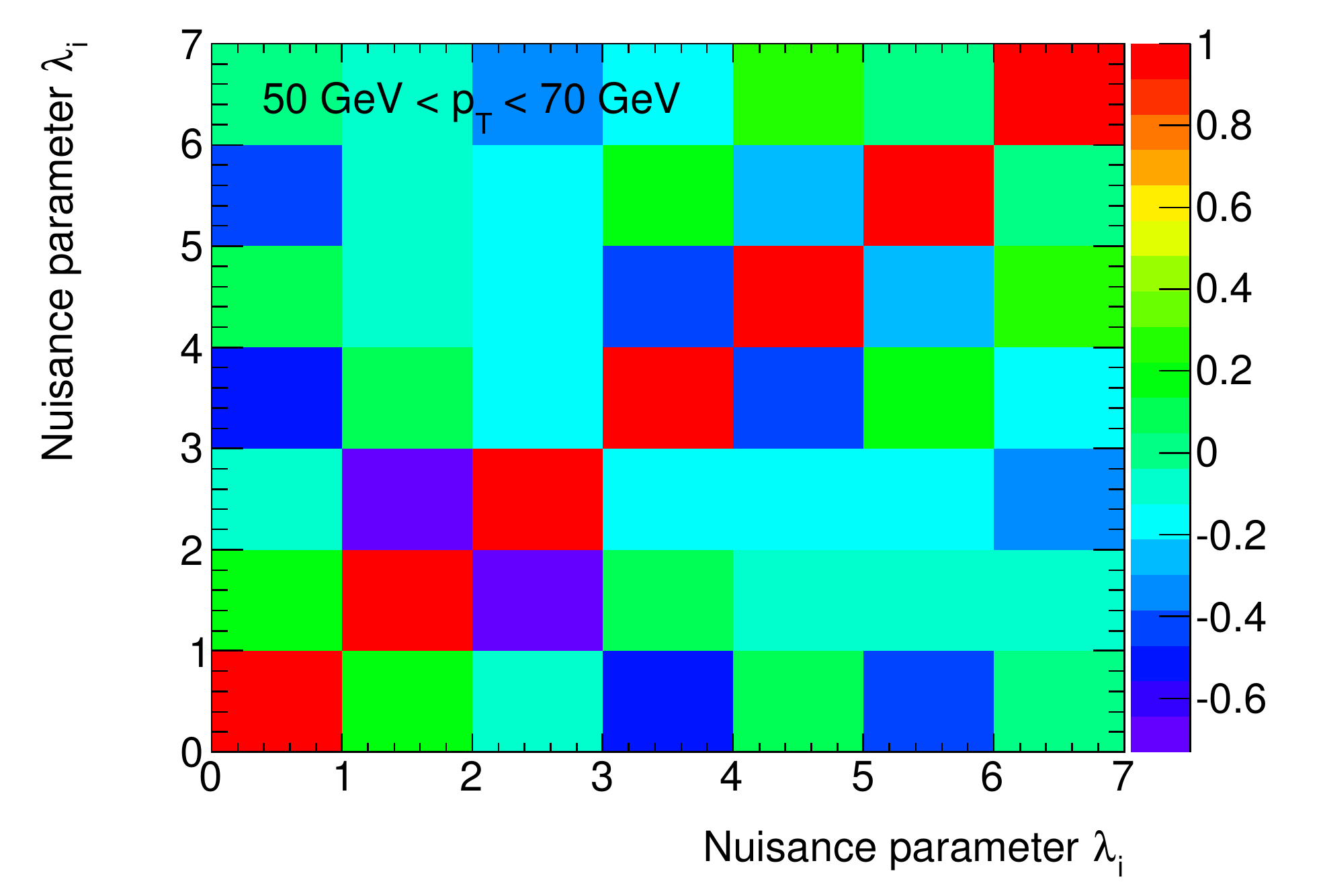}

\includegraphics[width=6.7cm,height=5.0cm]{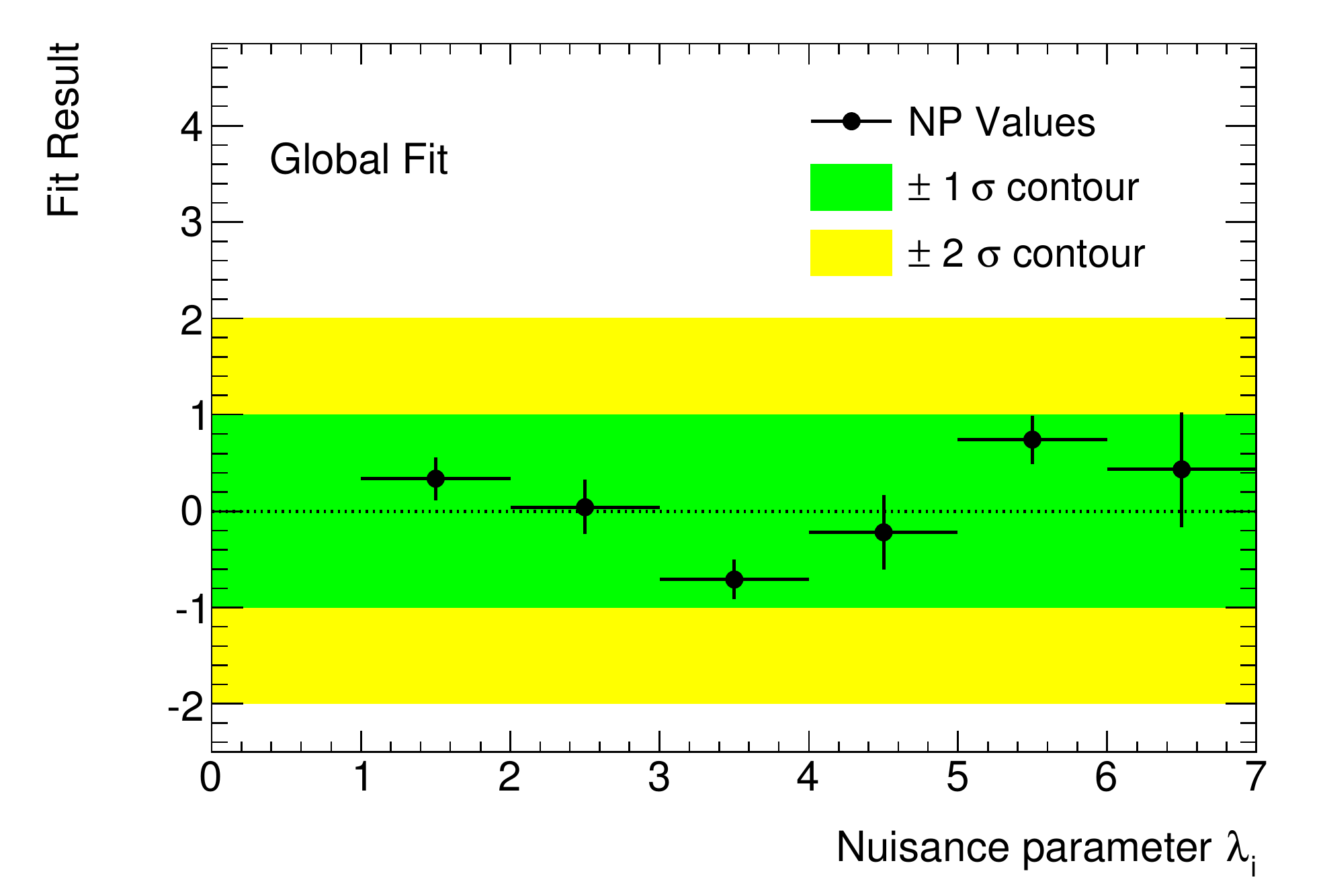}
\includegraphics[width=6.7cm,height=5.0cm]{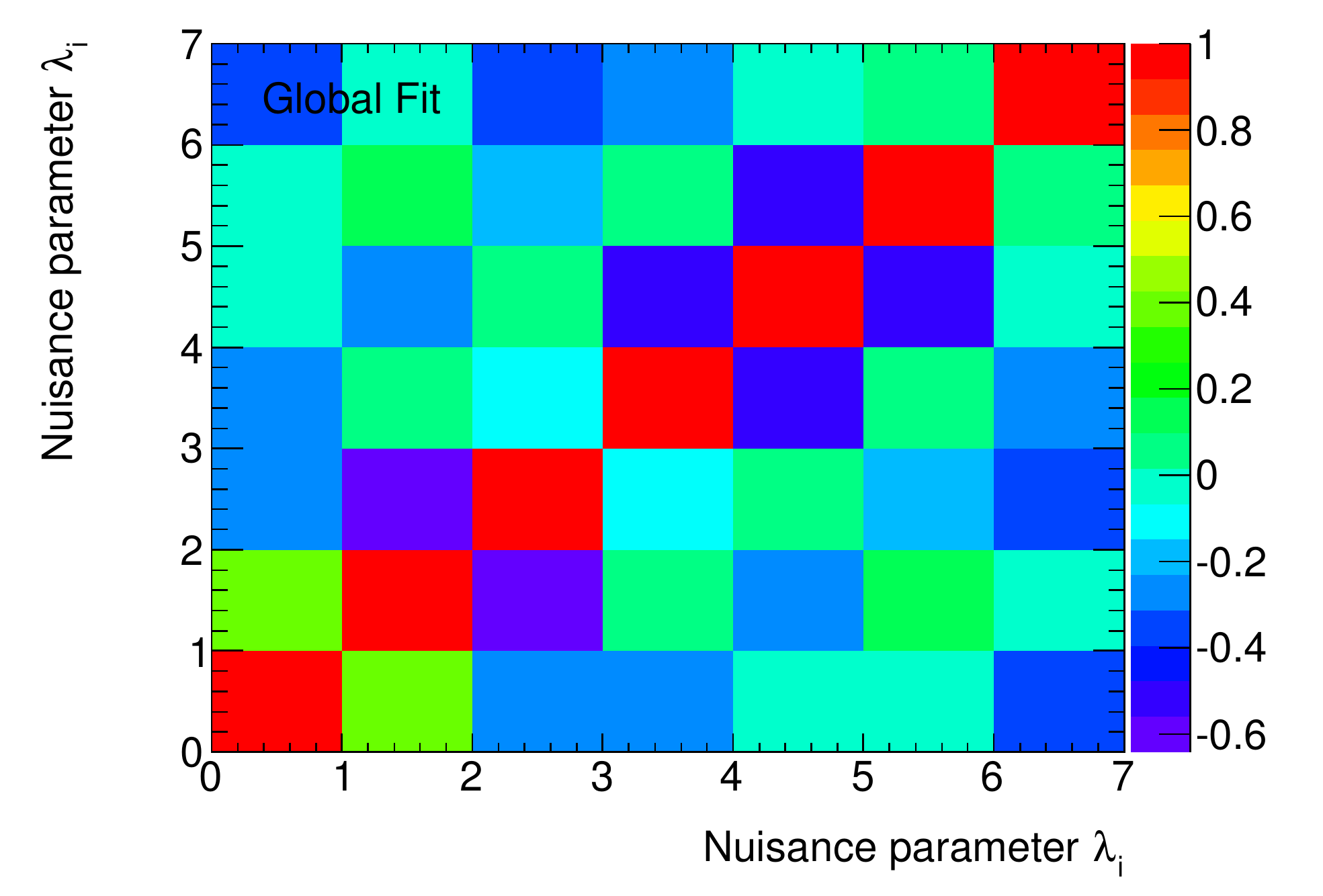}

\caption{Results for the nuisance parameters involved in the $\Lambda_s$ extraction (left column) and correlation matrices between them (right column) for each $\pt$ bin considered. The results obtained for the global fit are shown at the bottom row.}
\label{fig:fitLambda}
\end{figure}

In order to determine the QCD scale of the parton shower Monte Carlo which best fits the jet shape data, the dependence of the light-quark jet shapes on $\Lambda_s$ is studied. Figure \ref{fig:lambdaScan} shows the comparison of the light-jet shape data in \cite{atlasData} and the \textsc{Pythia} expectations for several values of $\Lambda_s$. The dependence of the jet shapes on $\Lambda_s$ is clearly seen from the figure.
In order to parametrise this dependence and obtain the interpolating functions $\phi_k(\Lambda_s)$ in Eq. \ref{eq:chi2B}, samples with $\Lambda_s$ varying from $20 \MeV$ to $300 \MeV$ in steps of $20 \MeV$ have been generated. To illustrate this dependence, Figure \ref{fig:lambda1} shows the points obtained from this scan together with the fitted functions $\phi_k(\Lambda_s)$ for $r = 0.02$ in each $\pt$ bin.
The fits using Eqs. \ref{eq:chi2A} and \ref{eq:chi2B} have been performed for every $\pt$ bin separately, and finally all of them are combined into a global fit to the three bins with $30 \GeV < \pt < 70 \GeV$. Fig. \ref{fig:fitLambda} shows the values of the nuisance parameters $\{\lambda_i\}$ involved in the fit, as well as the correlation matrix between them. The values of the nuisance parameters are always compatible with the $\pm 1\sigma$ band, fact which gives us confidence on the quality of the fit. The results of the fits to $\Lambda_s$ are summarised in Table \ref{tab:lambdaResults}, together with the fit uncertainties and the values of $\chi^2/N_{dof}$.

\begin{table}[H]
\caption{Summary of the results of the fit for $\Lambda_s$ using the light-jet shape data.}
\label{tab:lambdaResults}
\begin{center}
\begin{tabular}{cccc}
\hline
\T \B Bin & $\Lambda_s$ value (\MeV) & Fit error (\MeV) & $\chi^2/N_{dof}$\\
\hline
\T $30 \GeV < \pt < 40 \GeV$ & 187.5 & 24.0 & 10.6 / 9\\
\T $40 \GeV < \pt < 50 \GeV$ & 193.5 & 24.2 & 11.0 / 9\\
\T\B $50 \GeV < \pt < 70 \GeV$ & 137.7 & 17.3 & 7.8 / 9\\
\hline
\T \textbf{Global fit} & 162.1 & 9.6 & 39.0 / 29
\end{tabular}
\end{center}
\vspace{-0.3cm}
\end{table}
The nominal results obtained here have been derived using the one-loop solution to the renormalisation group equation (RGE) for the \textsc{Pythia} parton shower. In addition, the values of $\Lambda_s$ have also been extracted using \textsc{Herwig++} with the solutions to the RGE implemented up to two loops. The resulting value at one loop is $\Lambda_s = 160.7 \pm 15.3 \MeV$, in good agreement with the nominal value quoted above. For the two-loop case, the expression for the running strong coupling is \cite{pdg}
\newline
\begin{equation}
\alpha_s(Q^2) = \frac{1}{\beta_0\log x} \left[1- \frac{\beta_1}{\beta_0^2} \frac{\log\left(\log x \right)}{\log x}\right]; \ \ x = \frac{Q^2}{\Lambda^2}
\label{eq:asNLO}
\end{equation}
\newline
where $\beta_0$ is given in Eq. \ref{eq:asLO} and $\beta_1 = \frac{1}{(4\pi)^2}\left(102-\frac{38}{3}n_f\right)$. In this case, a value of $\Lambda_s = 276.1 \pm 17.3 \MeV$ is obtained, which is compatible with the value quoted in Ref. \cite{lambdaQCD} within the uncertainties. The effect of the two-loop running of the shower $\alpha_s$ on the $b$-quark mass will be explained in Sect. \ref{secUncLambda}.

\section{Determination of the $b$-quark mass $m_b$}
\label{secMb}
\begin{figure}[H]
\centering
\includegraphics[width=16.0cm,height=14.0cm]{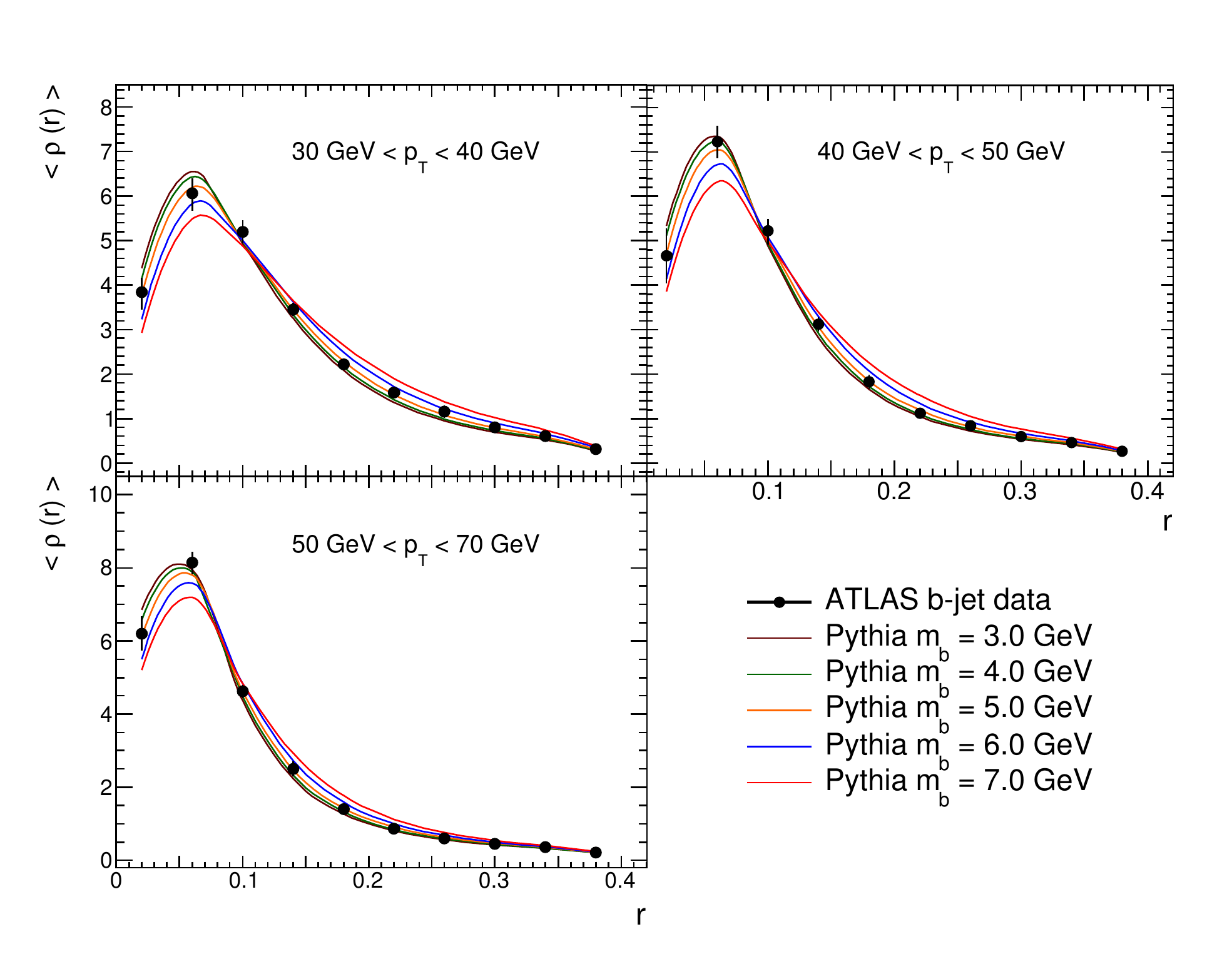}
\caption{Results of the $m_b$ scan compared to the ATLAS $b$-jet data in \cite{atlasData}. The QCD scale involved in the parton shower has been taken to be the one corresponding to the global fit to the light-jet data, $\Lambda_s = 162.1 \pm 9.6 \MeV$}
\label{fig:bMass}
\end{figure}

\begin{figure}[H]
\centering
\includegraphics[width=5.5cm,height=4.5cm]{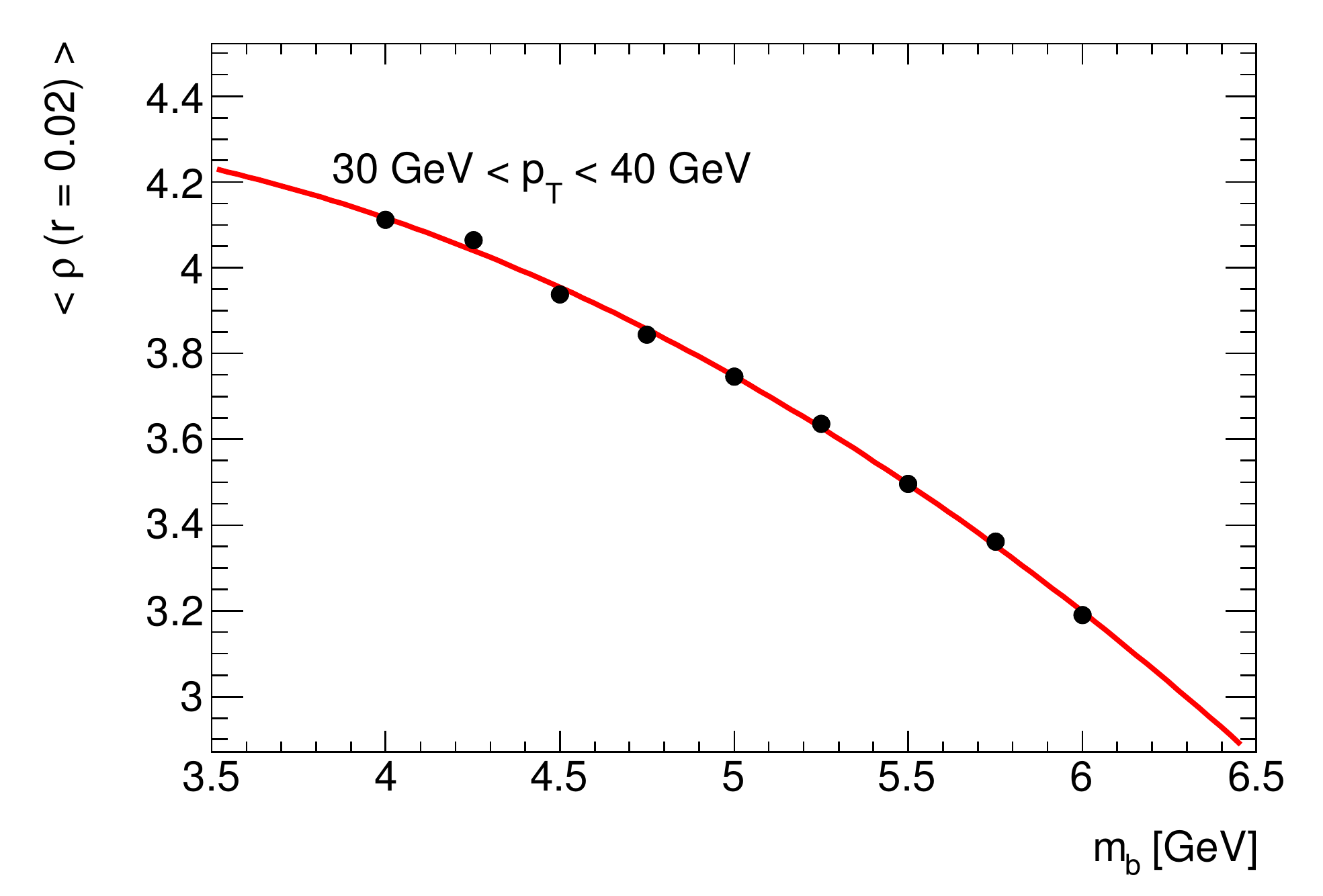}
\hspace{-0.25cm}
\includegraphics[width=5.5cm,height=4.5cm]{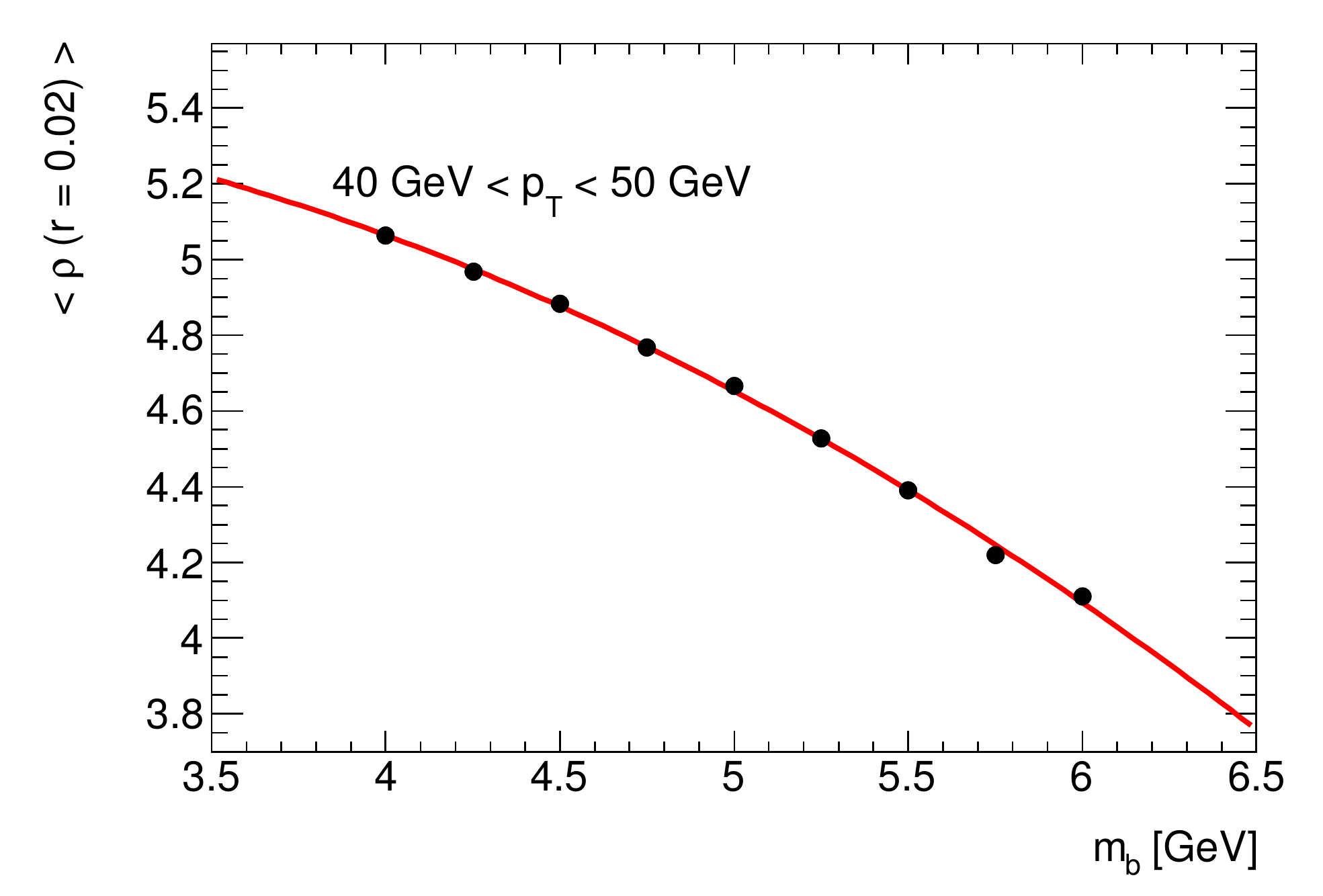}
\hspace{-0.25cm}
\includegraphics[width=5.5cm,height=4.5cm]{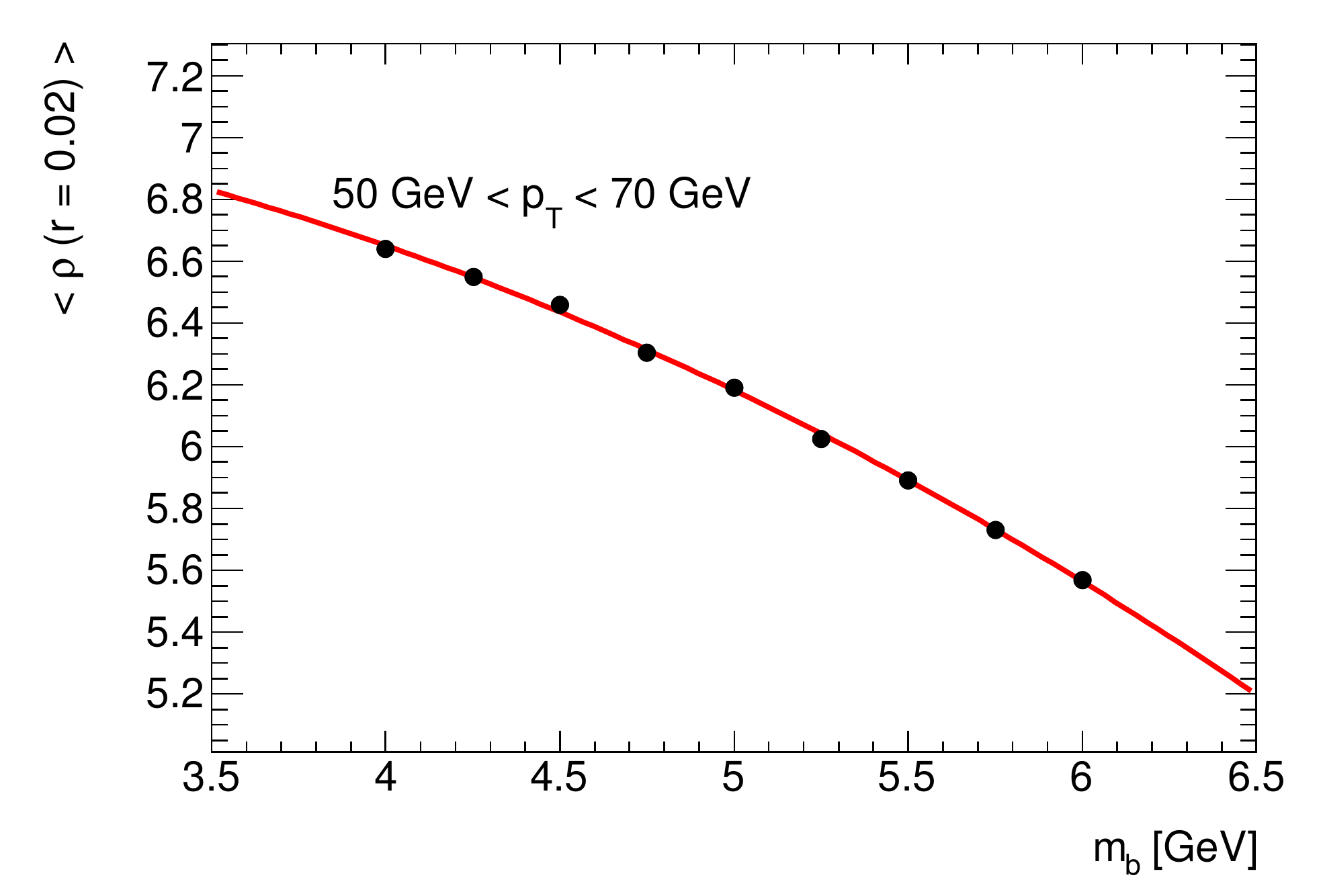}
\vspace{-0.7cm}
\caption{Dependence of the $b$-quark jet shape $\langle\rho(r = 0.02)\rangle$ with the $b$-quark mass $m_b$ for the $\pt$ intervals $30 \GeV < \pt < 40 \GeV$ (left), $40 \GeV < \pt < 50 \GeV$ (middle) and $50 \GeV < \pt < 70 \GeV$ (right), together with the interpolating functions $\phi_k(m_b)$.}
\label{fig:mb1}
\end{figure}

Once the parton shower scale $\Lambda_s$ is determined, the result is used to generate \textsc{Pythia} samples with several different values of the $b$-quark mass. The scan is performed in this case by varying this parameter from $4.0 \GeV$ to $6.0 \GeV$ in steps of $250 \MeV$. Figure \ref{fig:bMass} shows the comparison of the ATLAS $b$-jet shape data with the expectations from \textsc{Pythia} for several values of $m_b$, including those for $m_b = 3.0$ and $7.0 \GeV$. The value of the parton shower scale for the predictions shown in Fig. \ref{fig:bMass} is the one corresponding to the global fit to the light-jet shape data, namely $\Lambda_s = 162.1 \pm 9.6 \MeV$. The dependence of the jet shapes with the $b$-quark mass is more marked than in the case of $\Lambda_s$, providing high precision to the $m_b$ determination.
The $b$-jet shapes show a turn-over close to the jet cores due to the angular screening caused by the heavy mass of the $b$-quark. The larger $m_b$ is, the wider is the jet in the sense that the inner core has smaller energy deposits. The description of all bins provided by \textsc{Pythia} is excellent, showing that it is possible to perform a safe fit to the data. The parametrisation of the interpolating functions $\phi_k(m_b)$ describing the dependence of the differential $b$-jet shapes with the $b$-quark mass is done using second-order polynomials as in the case of the shower scale $\Lambda_s$. Figure \ref{fig:mb1} shows the dependence for $r = 0.02$ in each $\pt$ bin, as predicted by \textsc{Pythia}
The global fit has been performed including all $\pt$ bins and using the global value of the parton shower scale, $\Lambda_s = 162.1\pm 9.6 \MeV$. As a cross-check, for each $\pt$ bin, extra samples have been generated using the partial values of $\Lambda_s$ shown in Table \ref{tab:lambdaResults}. The agreement between all the extracted values of $m_b$ is excellent, as can be seen in Table \ref{tab:mbResults}.

\begin{table}[H]
\caption{Summary of the results of the fits for $m_b$ using the $b$-jet shape data and the corresponding value of $\Lambda_s$ for each bin listed in table \ref{tab:lambdaResults}. The global fit is performed using the globally extracted value of the parton shower scale $\Lambda_s = 162.1 \MeV$.}
\label{tab:mbResults}
\begin{center}
\begin{tabular}{cccc}
\hline
\T \B Bin & $m_b$ value (\GeV) & Fit error (\GeV) & $\chi^2/N_{dof}$\\
\hline
\T $30 \GeV < \pt < 40 \GeV$ & 5.00 & 0.14 & 8.28 / 9\\
\T $40 \GeV < \pt < 50 \GeV$ & 4.82 & 0.19 & 10.41 / 9\\
\T\B $50 \GeV < \pt < 70 \GeV$ & 4.82 & 0.13 & 11.99 / 9\\
\hline
\T \textbf{Global fit} & 4.86 & 0.08 & 43.04 / 29\\
\end{tabular}
\end{center}
\end{table}

\begin{figure}[H]
\centering
\includegraphics[width=6.7cm,height=5.0cm]{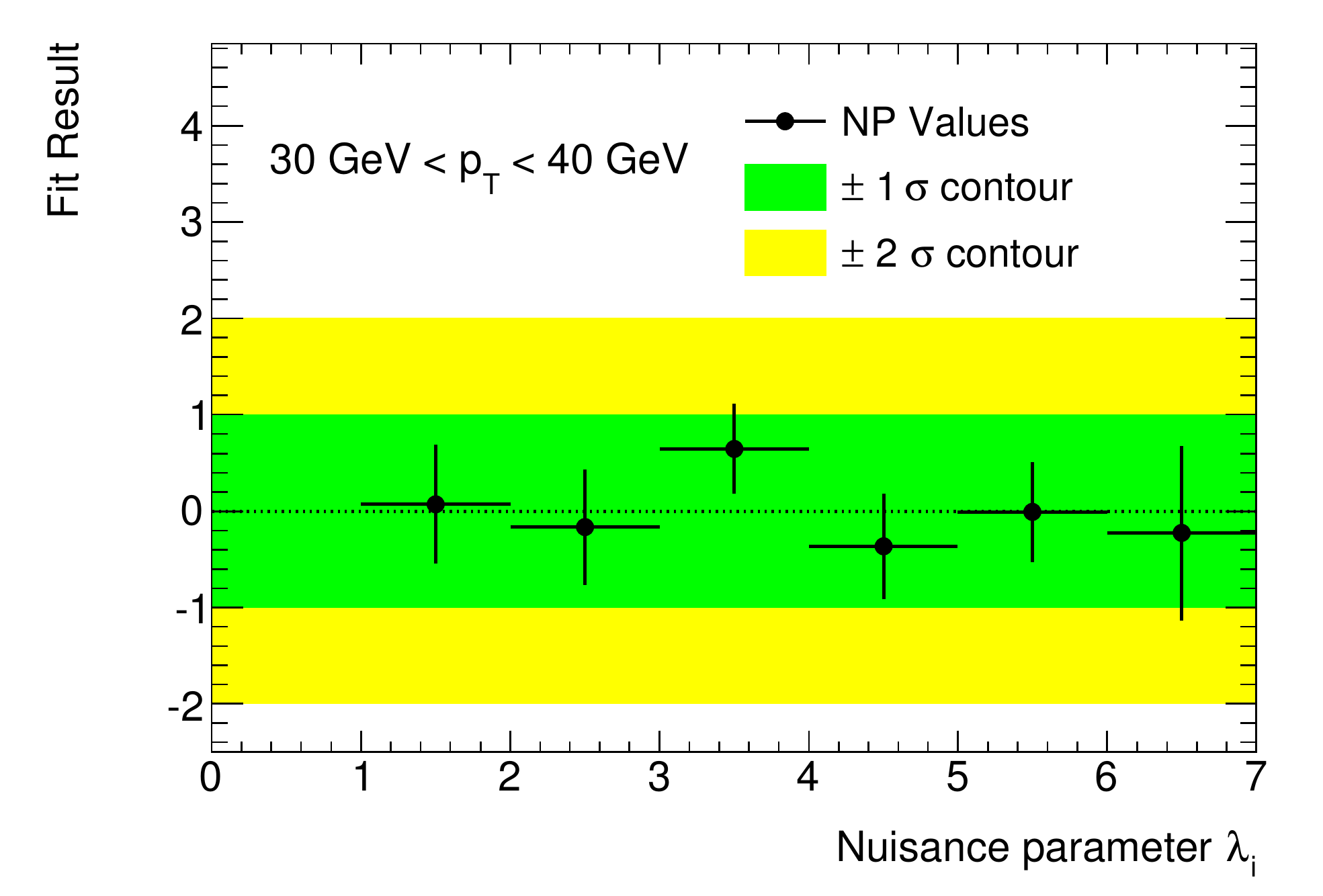}
\includegraphics[width=6.7cm,height=5.0cm]{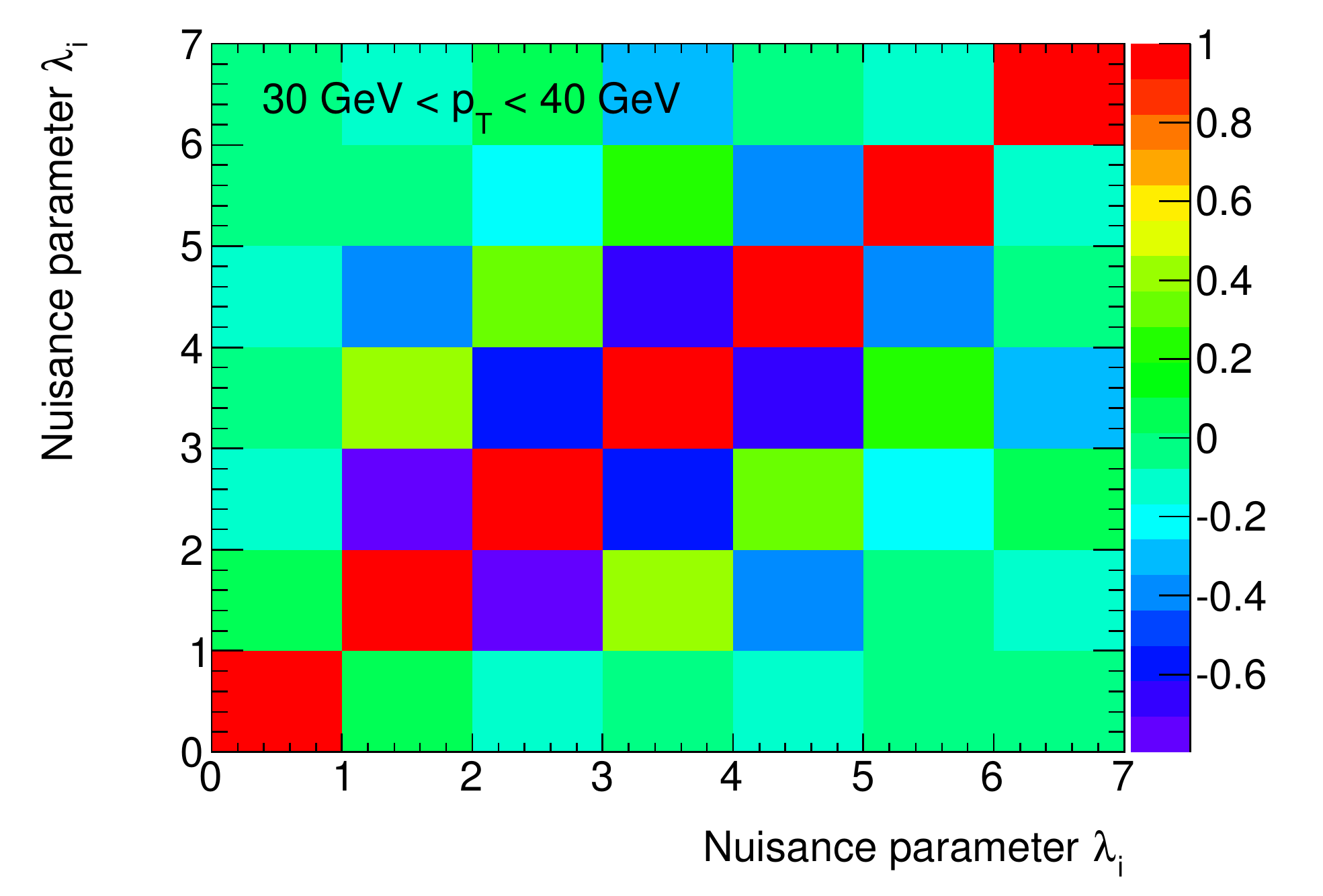}

\includegraphics[width=6.7cm,height=5.0cm]{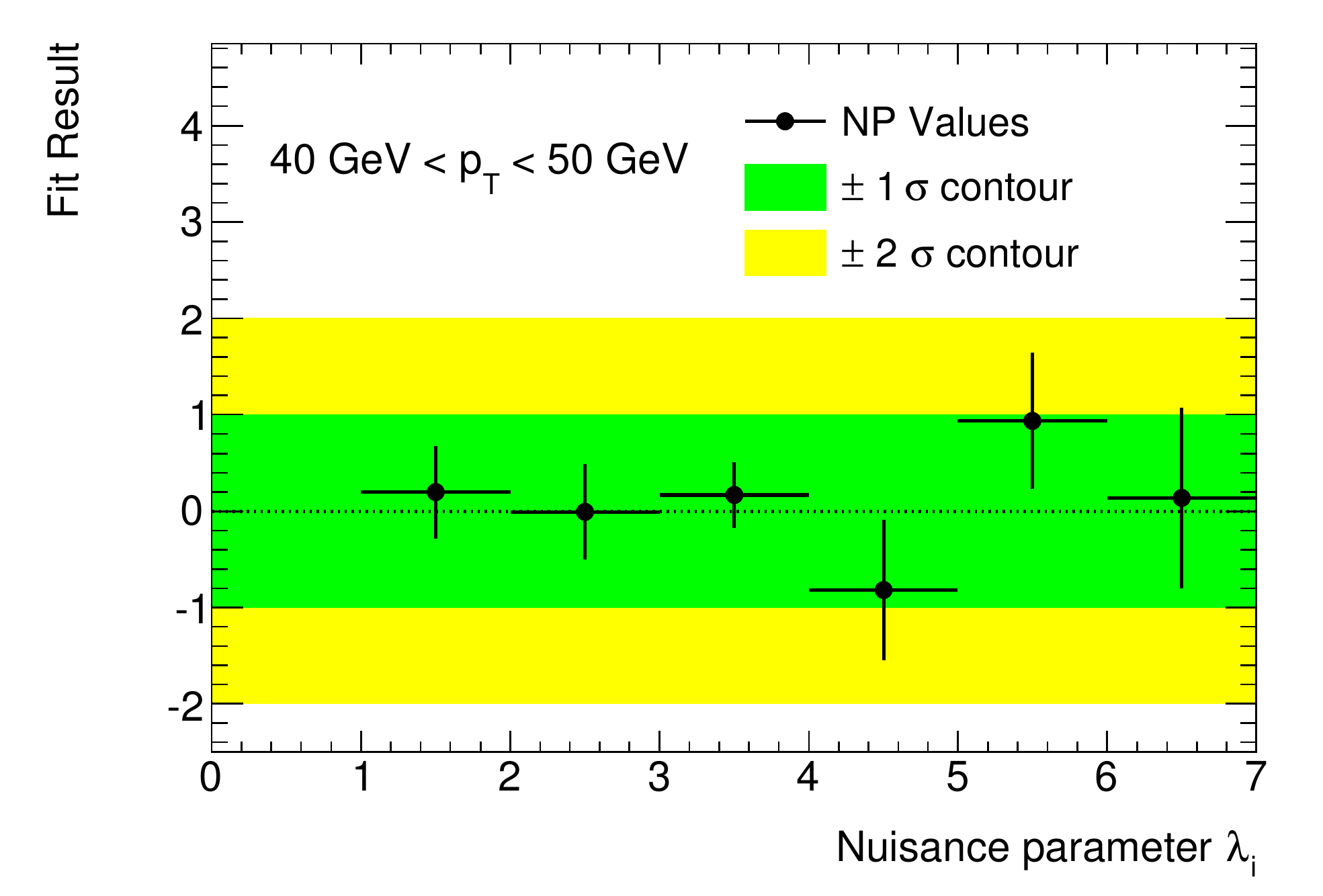}
\includegraphics[width=6.7cm,height=5.0cm]{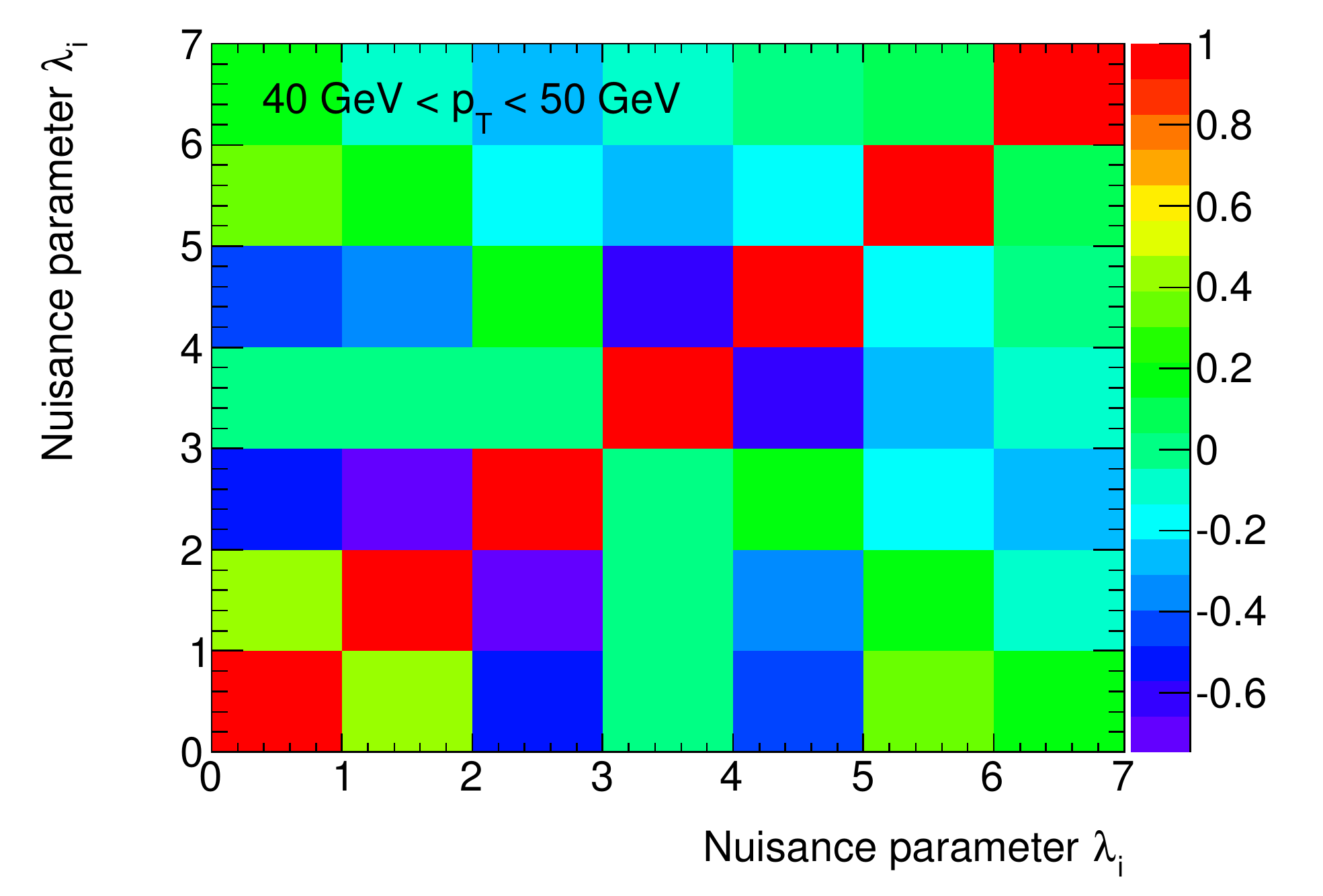}

\includegraphics[width=6.7cm,height=5.0cm]{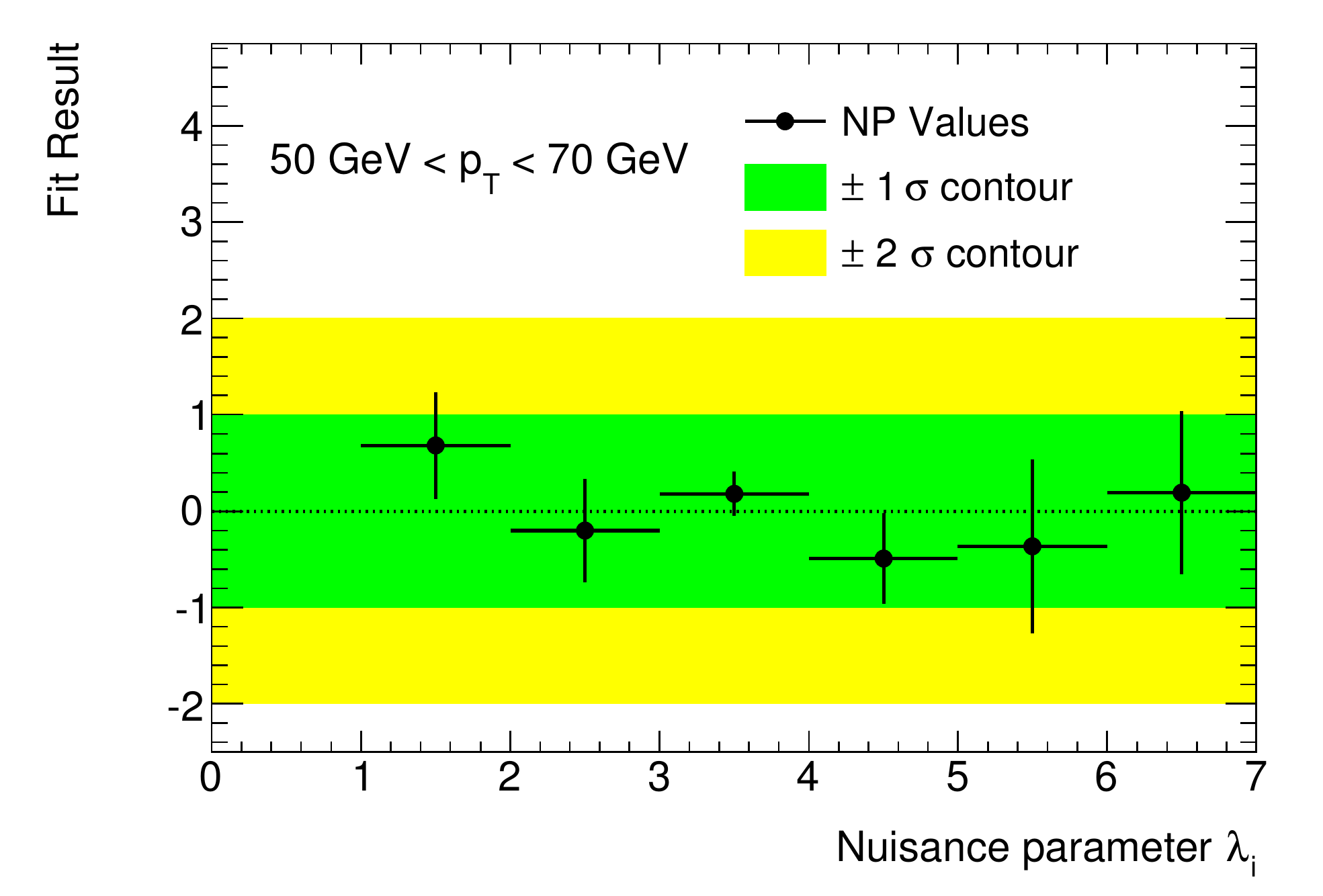}
\includegraphics[width=6.7cm,height=5.0cm]{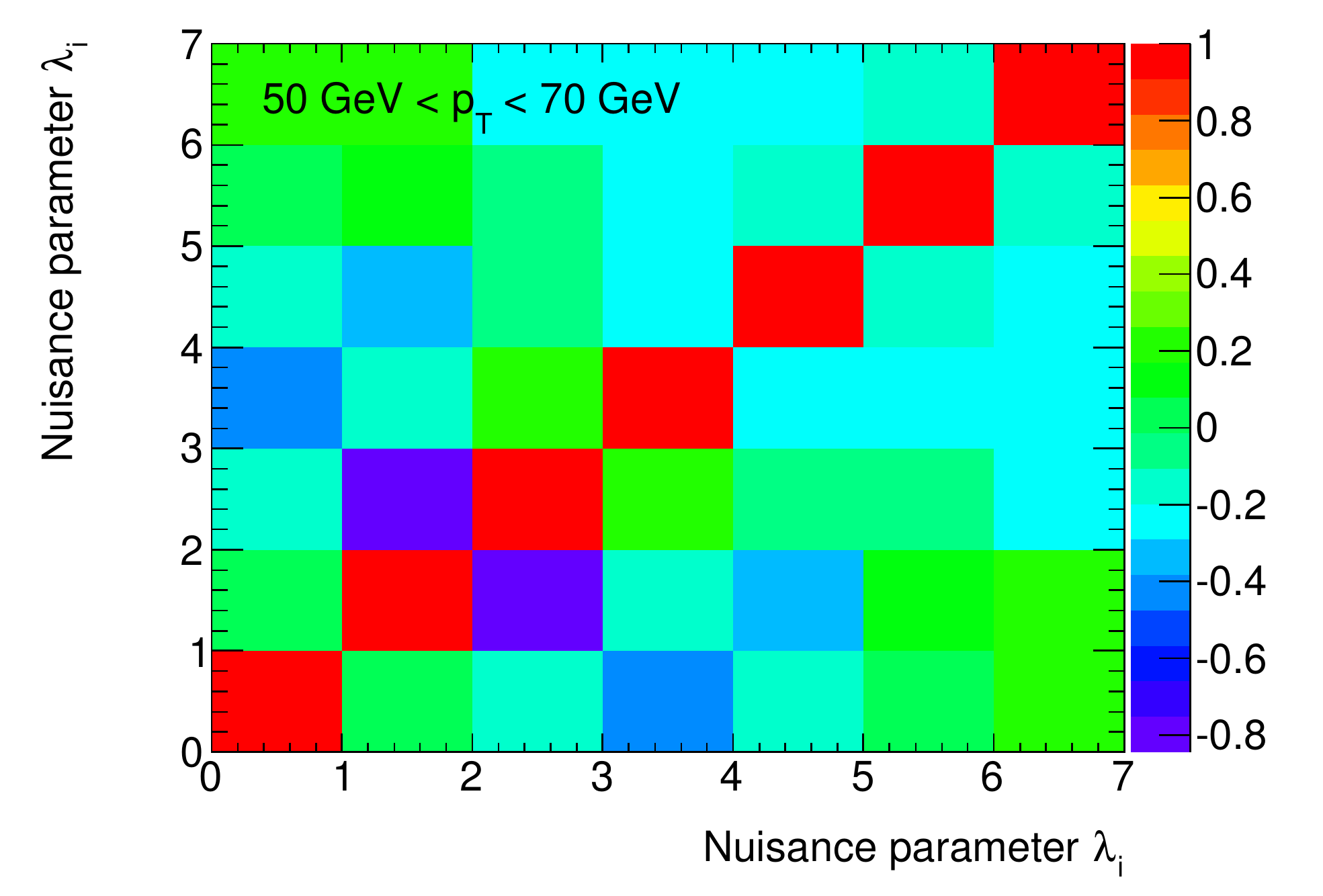}

\includegraphics[width=6.7cm,height=5.0cm]{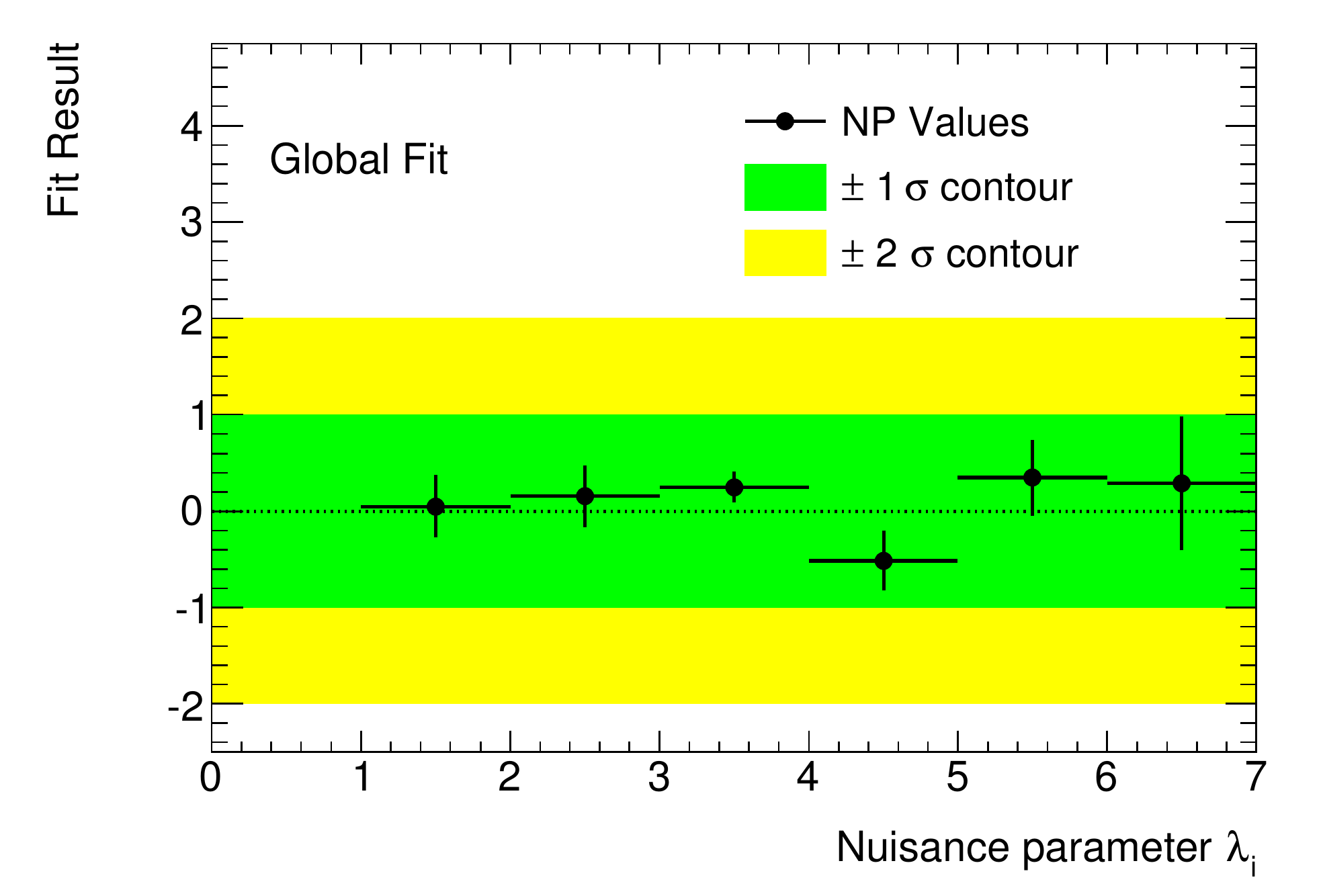}
\includegraphics[width=6.7cm,height=5.0cm]{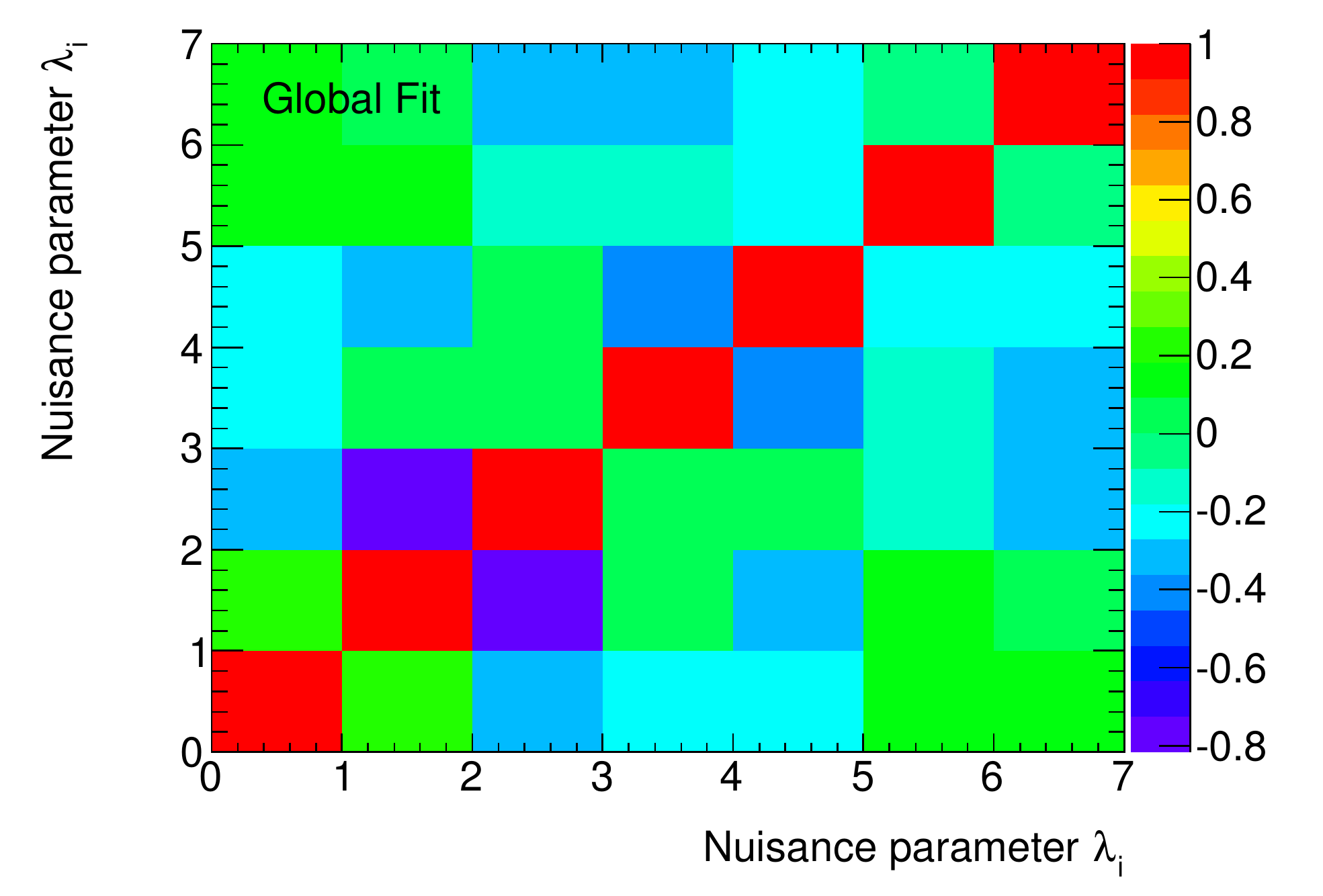}

\caption{Results for the nuisance parameters involved in the $m_b$ extraction (left column) and correlation matrices between them (right column) for each $\pt$ bin considered. The results obtained for the global fit are shown at the bottom row. For each $\pt$ bin, the corresponding value of $\Lambda_s$ listed in Table \ref{tab:lambdaResults} has been used. For the global fit, the globally extracted value is $\Lambda_s = 162.1 \MeV$.}
\label{fig:fitMb}
\end{figure}

As before, the values of the nuisance parameters and the correlation matrices between them are shown in Figure \ref{fig:fitMb} for the fits performed using each extracted value of the shower scale. As in the previous case, we find that the nuisance parameters are well behaved, being always compatible with the $\pm 1\sigma$ contour band. This is specially important for the global fit, as its result will be taken as the central value for our determination. As can be seen in the lower part of Figure \ref{fig:fitMb}, the behaviour of the fit parameters is very good.

\section{Theoretical uncertainties}
\label{secUnc}
In this section, the uncertainties on the theory are discussed. They come from several sources, including the modelling of the parton shower, hadronisation and multiple parton interactions. Other effects such as the amount of initial and final-state radiation, the colour reconnection model and the error on the determination of the parton shower scale $\Lambda_s$ are also studied. The generator modelling uncertainty is the main source of uncertainty for these analysis, not being greater than $400 \MeV$ in terms of the extracted $b$-quark mass. All variations of the theoretical distributions are performed with respect to the nominal sample, produced using the fitted values of $\Lambda_s$ and $m_b$.
\subsection{Generator modelling}
\begin{figure}[H]
\centering
\includegraphics[width=5.5cm,height=5.9cm]{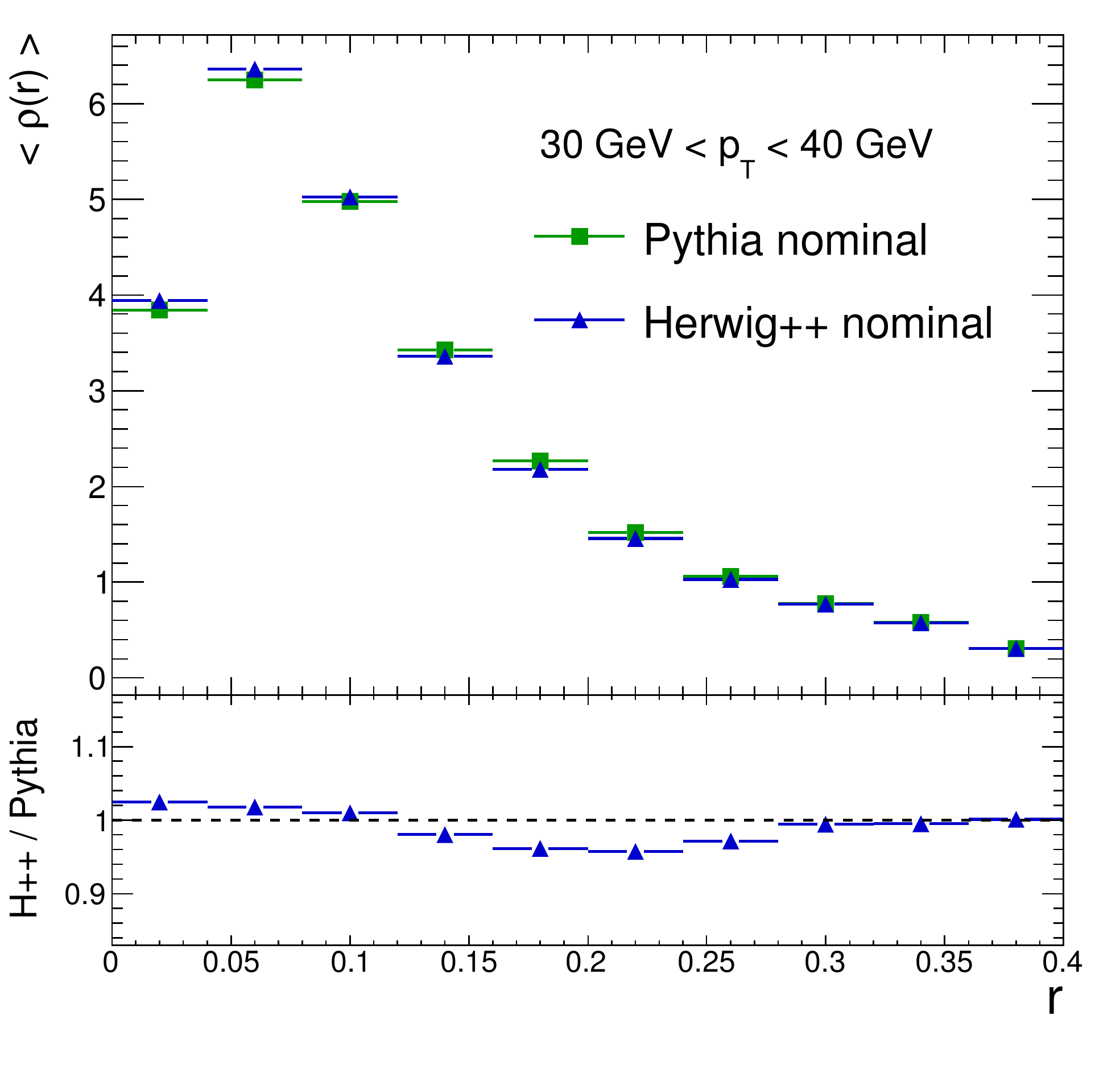}
\hspace{-0.25cm}
\includegraphics[width=5.5cm,height=5.9cm]{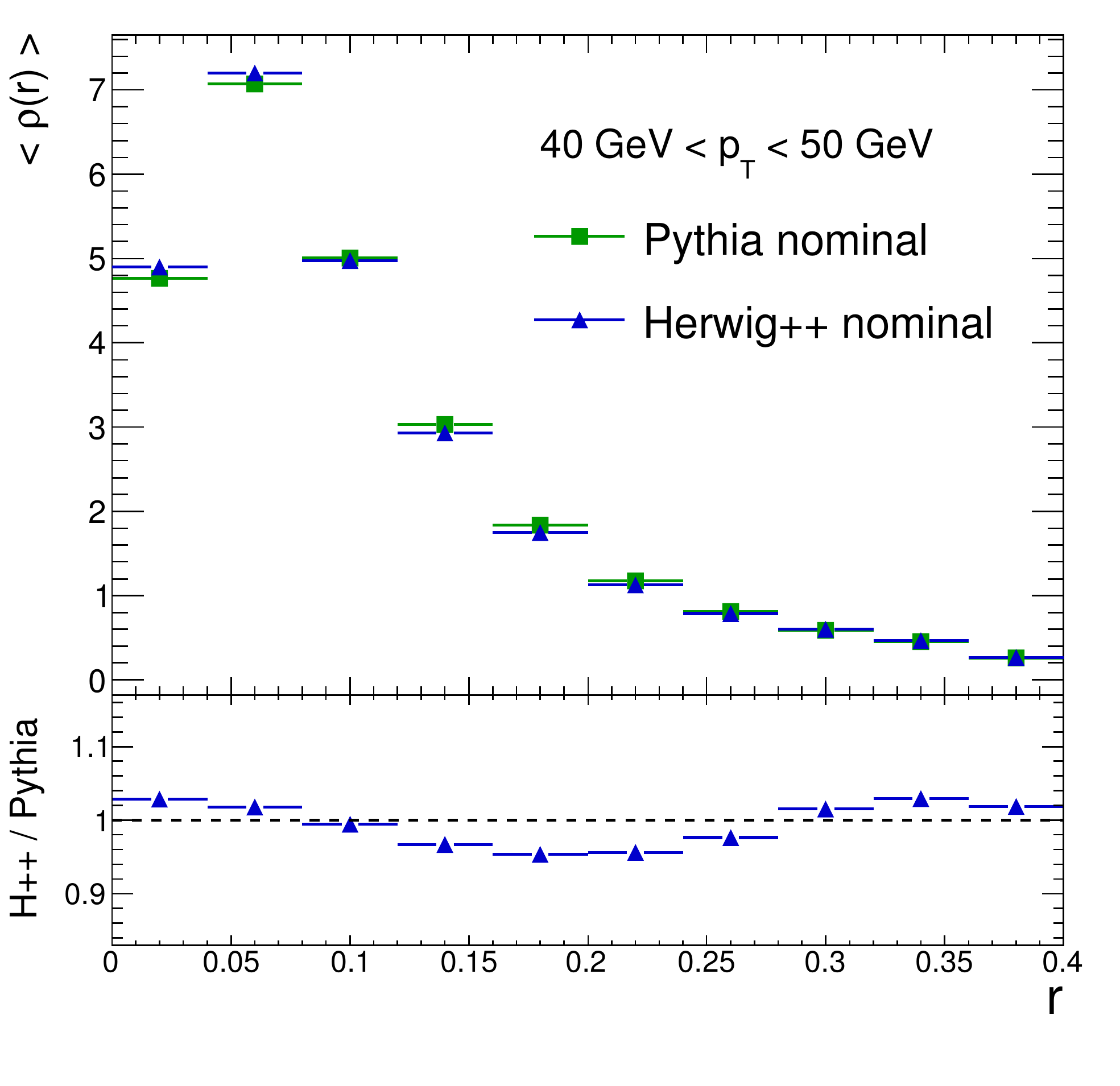}
\hspace{-0.25cm}
\includegraphics[width=5.5cm,height=5.9cm]{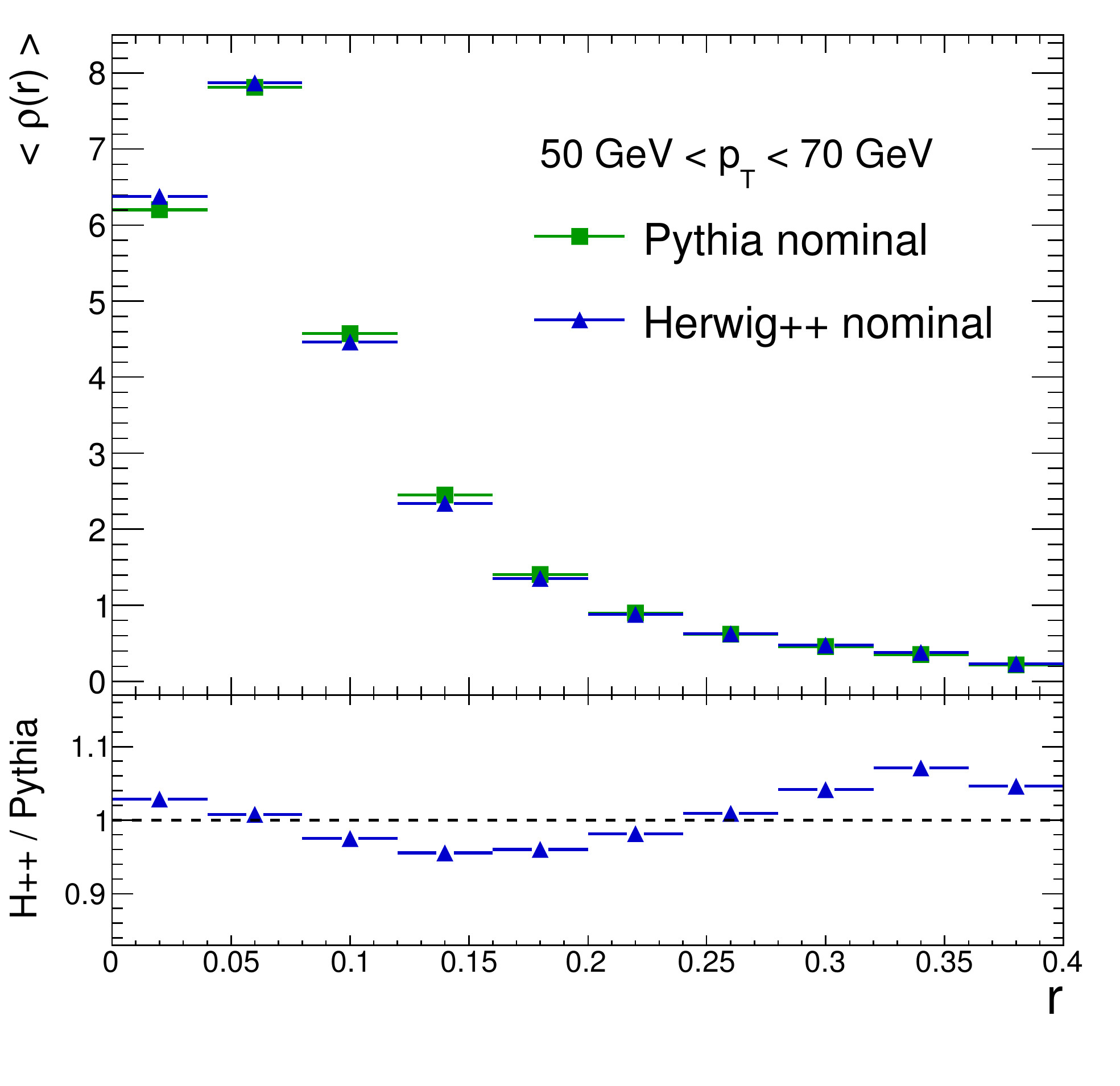}
\vspace{-1.cm}
\caption{The difference between the \textsc{Pythia} and \textsc{Herwig++} predictions for the fitted values $m_b = 4.86 \GeV$ and $\Lambda_s = 162.1 \MeV$. This difference, arising from the parton shower and hadronisation, is the source of the theoretical uncertainty due to the generator modelling.}
\label{fig:hadronUnc}
\end{figure}

The \textsc{Pythia} predictions use virtuality-ordered parton showers and the Lund string model for the hadronisation. In order to study the impact of this choice on the extraction of $m_b$, a sample of $t\bar{t}$ events has been generated using the \textsc{Herwig++} Monte Carlo program \cite{herwigpp}, which incorporates angular-ordered parton showers as well as the cluster hadronisation model. The modelling of the underlying event (multiparton interactions) is also different between both approaches. For \textsc{Herwig++}, the \textsc{LHC-UE7-2} tune has been chosen. This is based on the ATLAS measurements of the underlying event using charged particles \cite{atlasUE}. On the other hand, \textsc{Pythia} uses the so-called \textsc{Tune A} as default \cite{tuneA}, which is based on the correct description of many Tevatron measurements.\\
\newline
In Figure \ref{fig:hadronUnc}, the nominal prediction by \textsc{Pythia} is compared to the nominal predictions by \textsc{Herwig++}.
In order to study the impact of these differences on the determination of $m_b$, the full analysis has been repeated using \textsc{Herwig++}. In this case, the $b$-quark mass is scanned by varying both the \textsc{NominalMass} and the \textsc{ConstituentMass} flags for \textsc{/Herwig/Particles/b} and \textsc{/Herwig/Particles/bbar}. The value of $\Lambda_s$ has been set to $160.7 \pm 15.3 \MeV$, which is the value obtained in Section \ref{secLambda} for the \textsc{Herwig++} approach. The result for the $b$-quark mass is $m_b = 5.25 \pm 0.09 \GeV$, and the difference with respect to the nominal value is symmetrised and ascribed as a theoretical uncertainty.
\subsection{Initial-state radiation}
\begin{figure}[H]
\centering
\includegraphics[width=5.5cm,height=5.9cm]{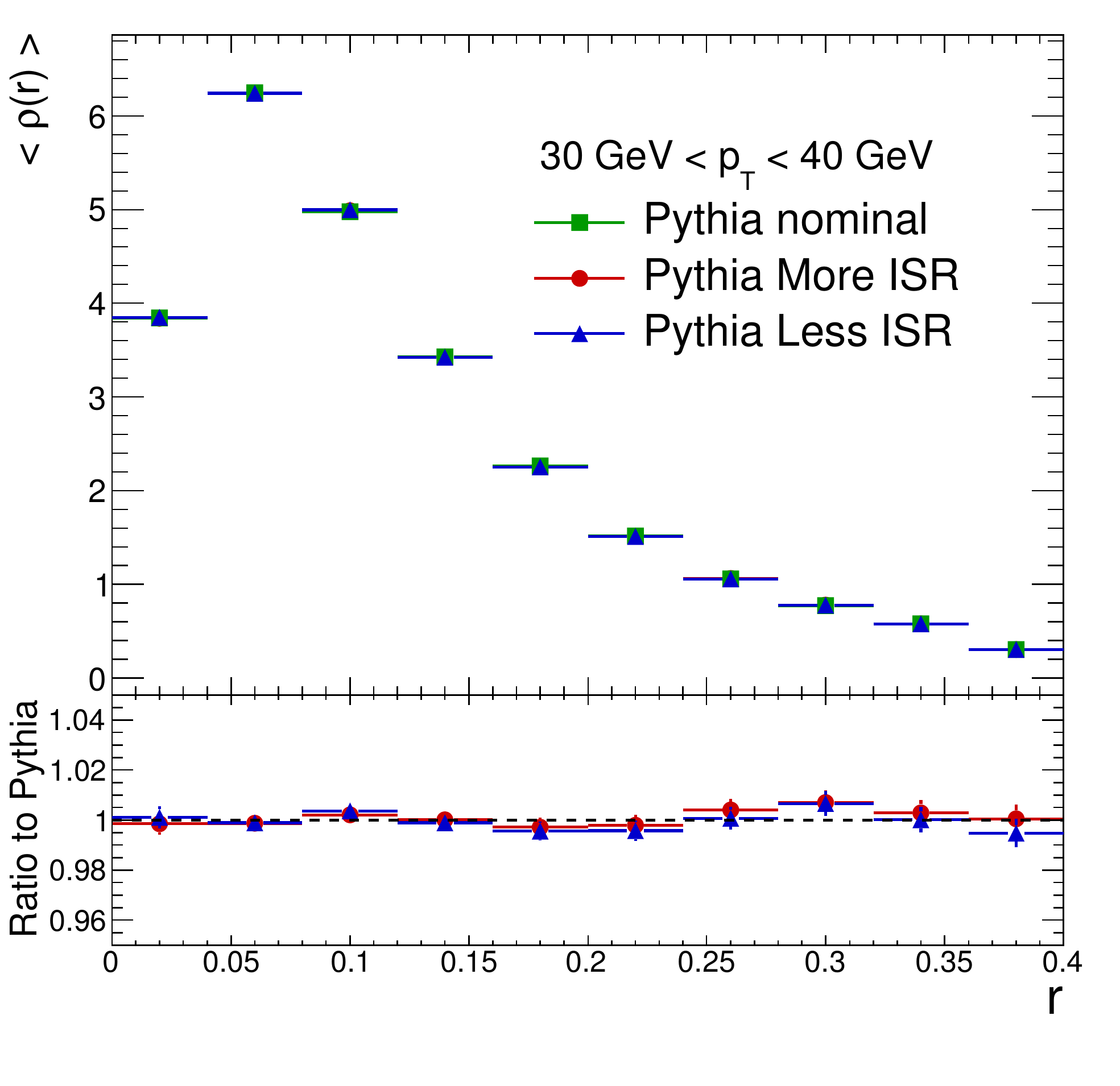}
\hspace{-0.25cm}
\includegraphics[width=5.5cm,height=5.9cm]{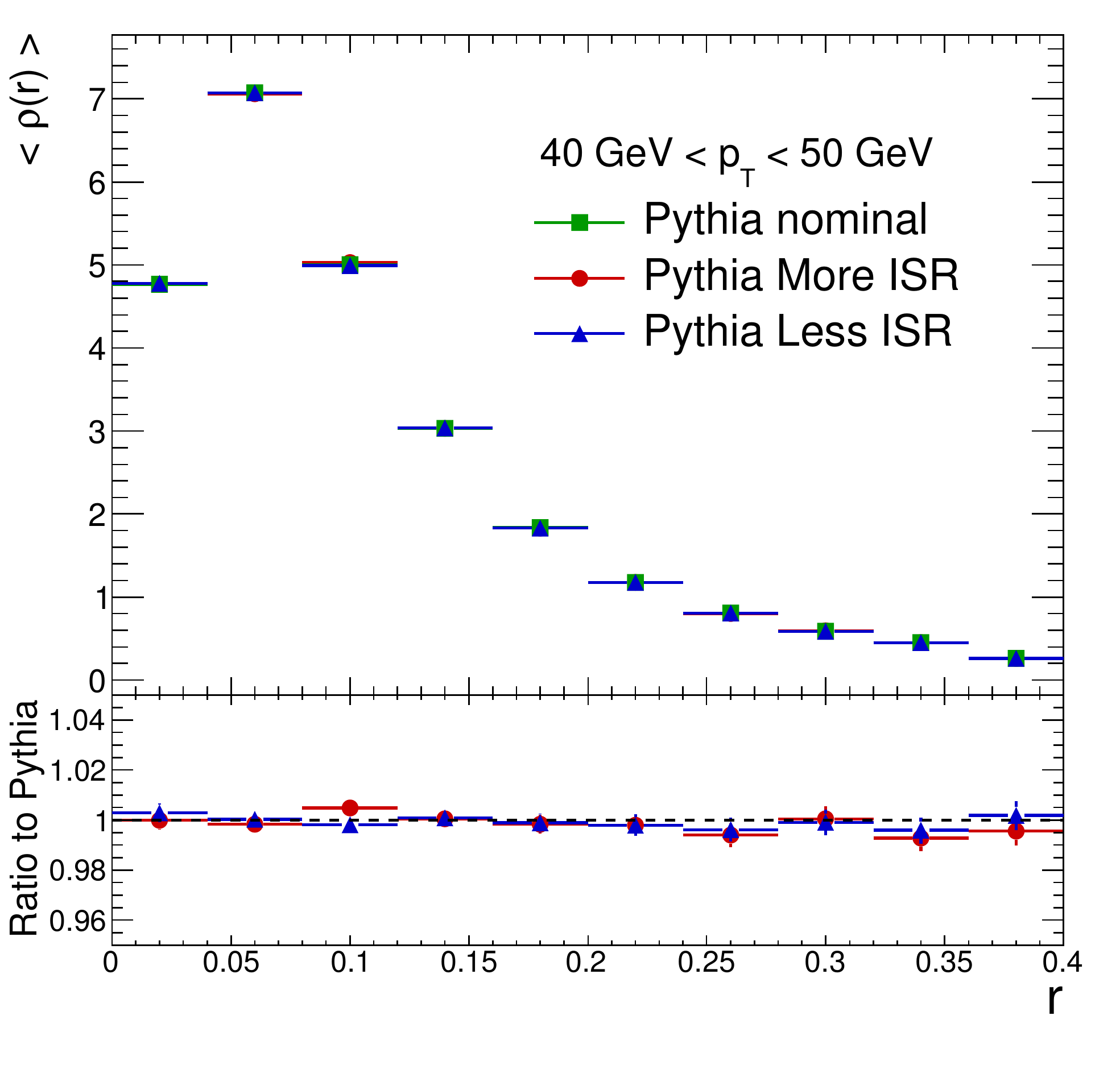}
\hspace{-0.25cm}
\includegraphics[width=5.5cm,height=5.9cm]{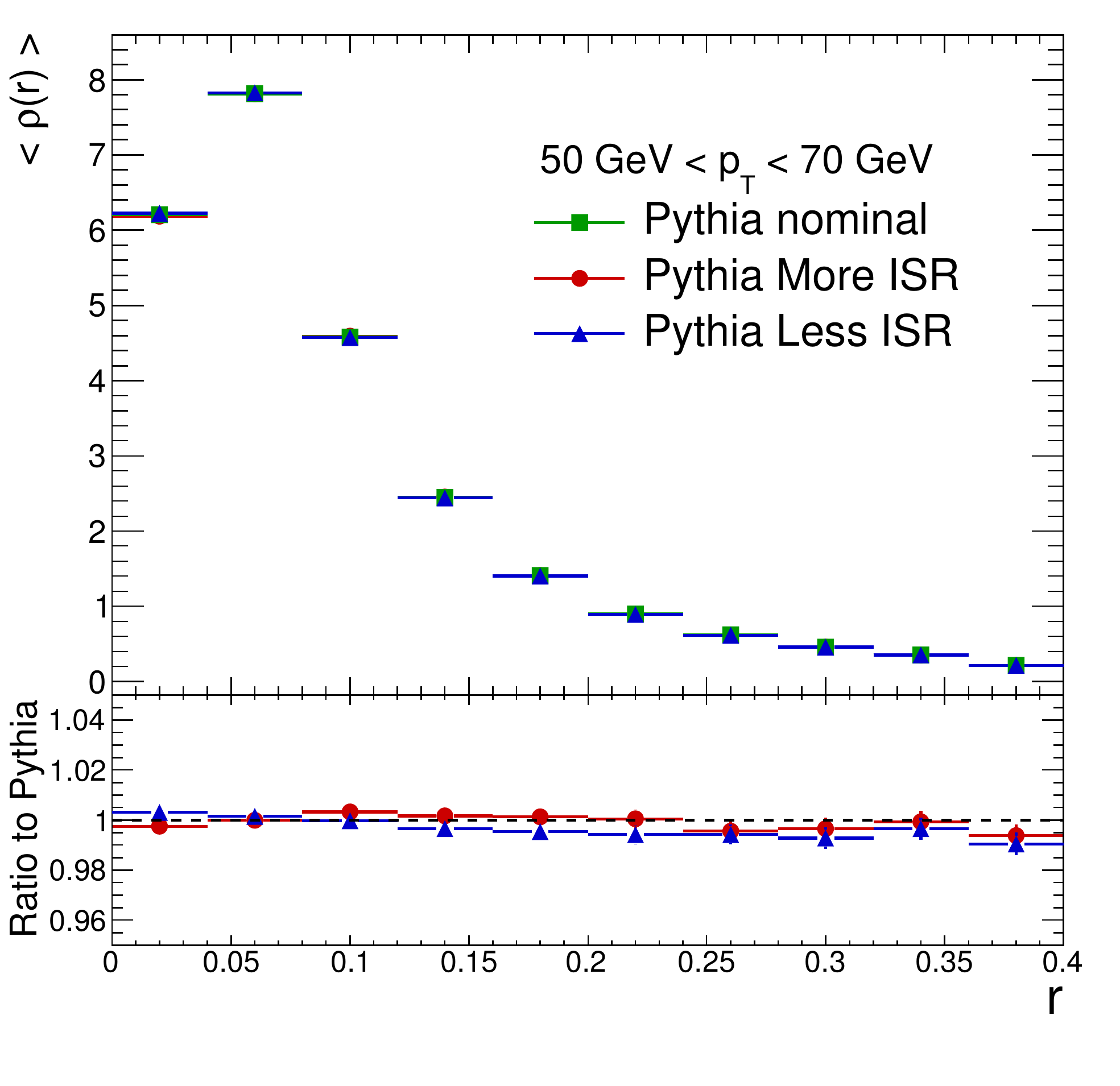}
\vspace{-1.cm}
\caption{The effects of the initial-state radiation on the $b$-jet shapes.}
\label{fig:isr}
\end{figure}

The amount of initial-state radiation (ISR) can lead to differences in the jet shapes. To test this effect, two additional samples with reduced and enhanced levels of ISR are generated. The ISR is controlled in \textsc{Pythia} using the parameters \textsc{PARP(67)} and \textsc{PARP(64)}. To decrease the ISR, the parameters are set to 0.5 and 4.0 respectively. To increase ISR, they are set to 6.0 and 0.25, respectively. These specifications have been widely used in several ATLAS analyses such as the study of $t\bar{t}$ production with a veto on central jet activity \cite{gapFraction}. The effects of these changes on the prediction and the comparison to the nominal \textsc{Pythia} sample are shown in Fig. \ref{fig:isr}. The effect of these variations on the $b$-quark mass is around $20 \MeV$, which is negligible for the final result, compared to the generator uncertainty.

\subsection{Final-state radiation}
\begin{figure}[H]
\centering
\includegraphics[width=5.5cm,height=5.9cm]{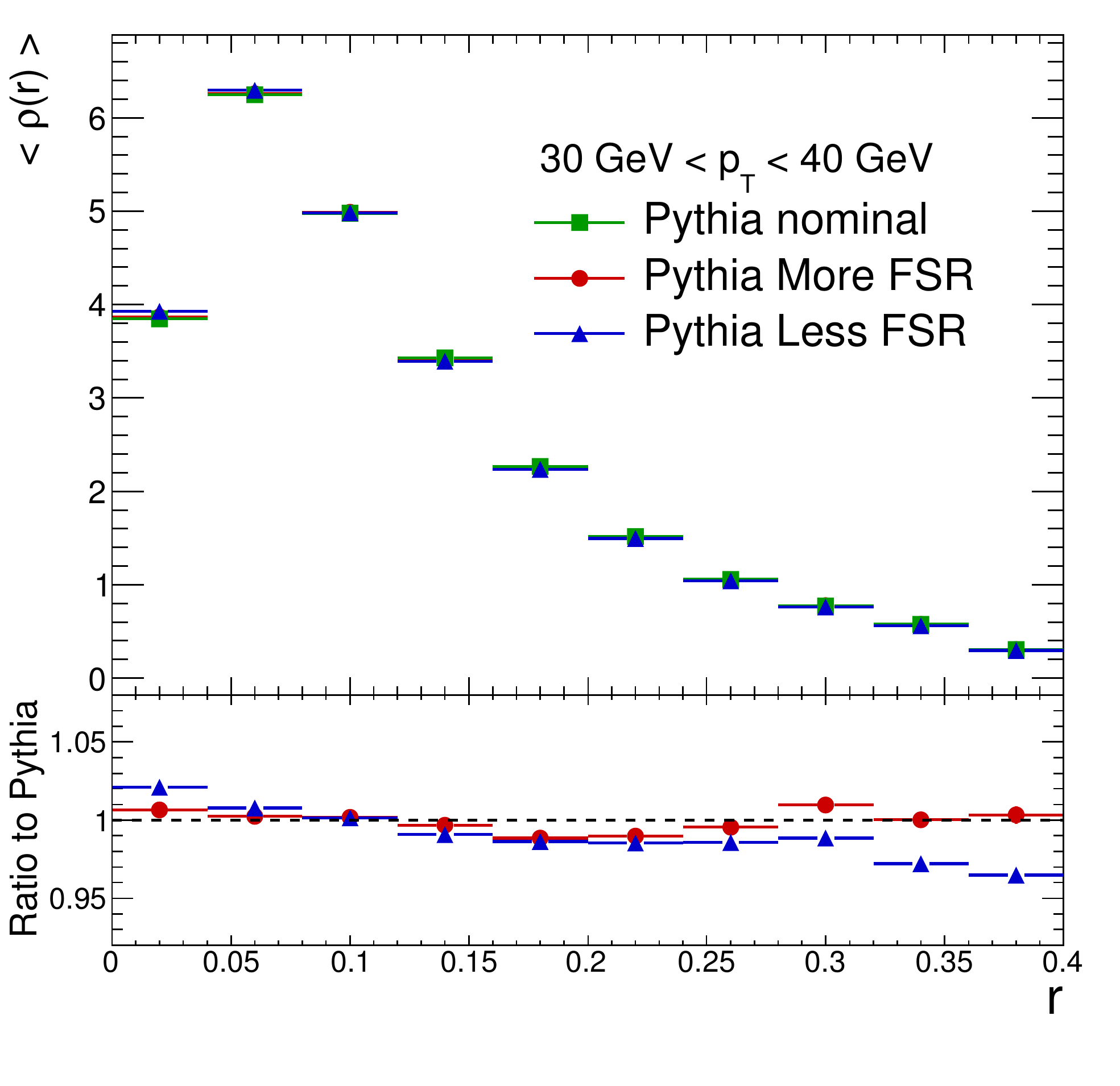}
\hspace{-0.25cm}
\includegraphics[width=5.5cm,height=5.9cm]{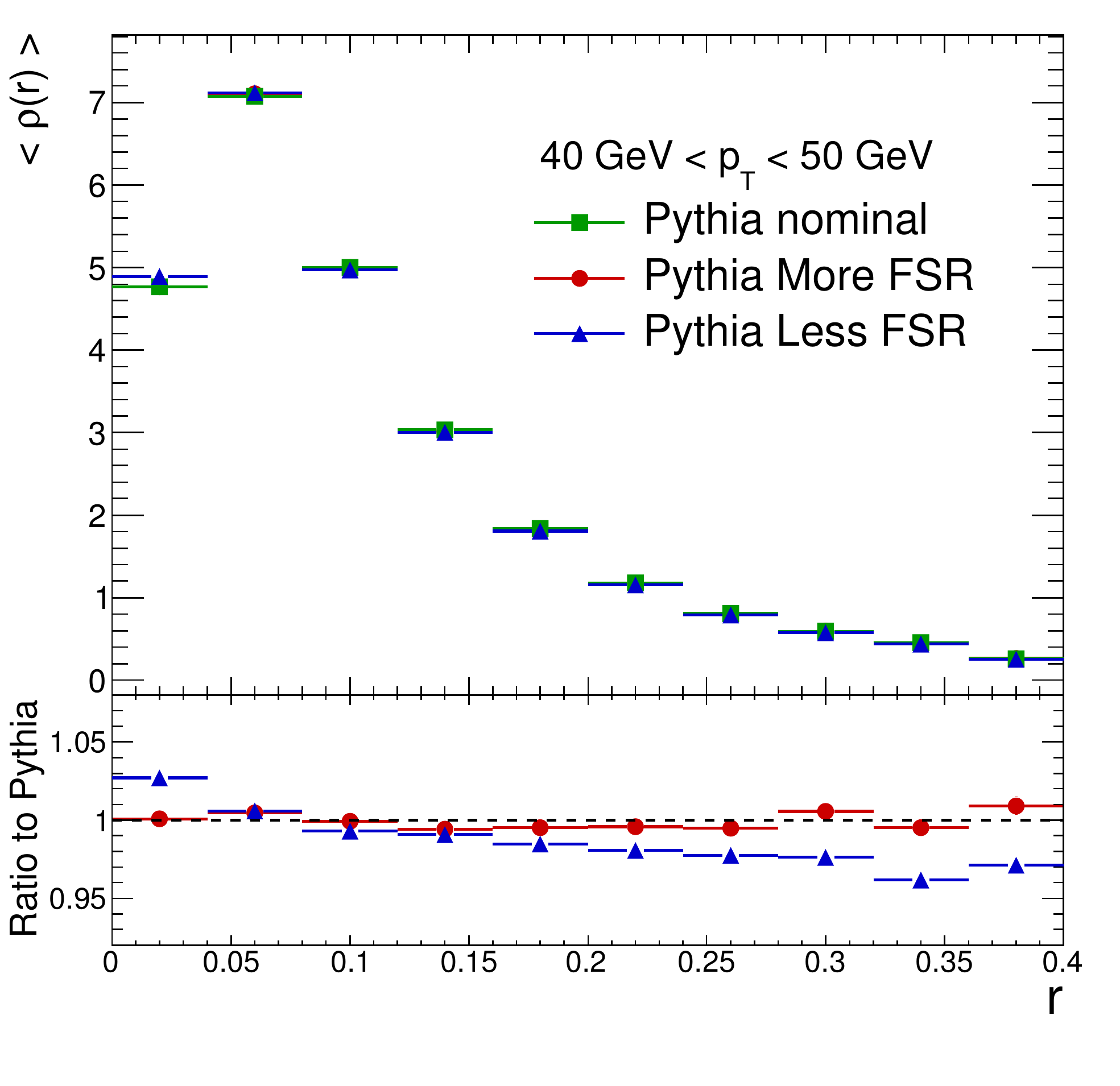}
\hspace{-0.25cm}
\includegraphics[width=5.5cm,height=5.9cm]{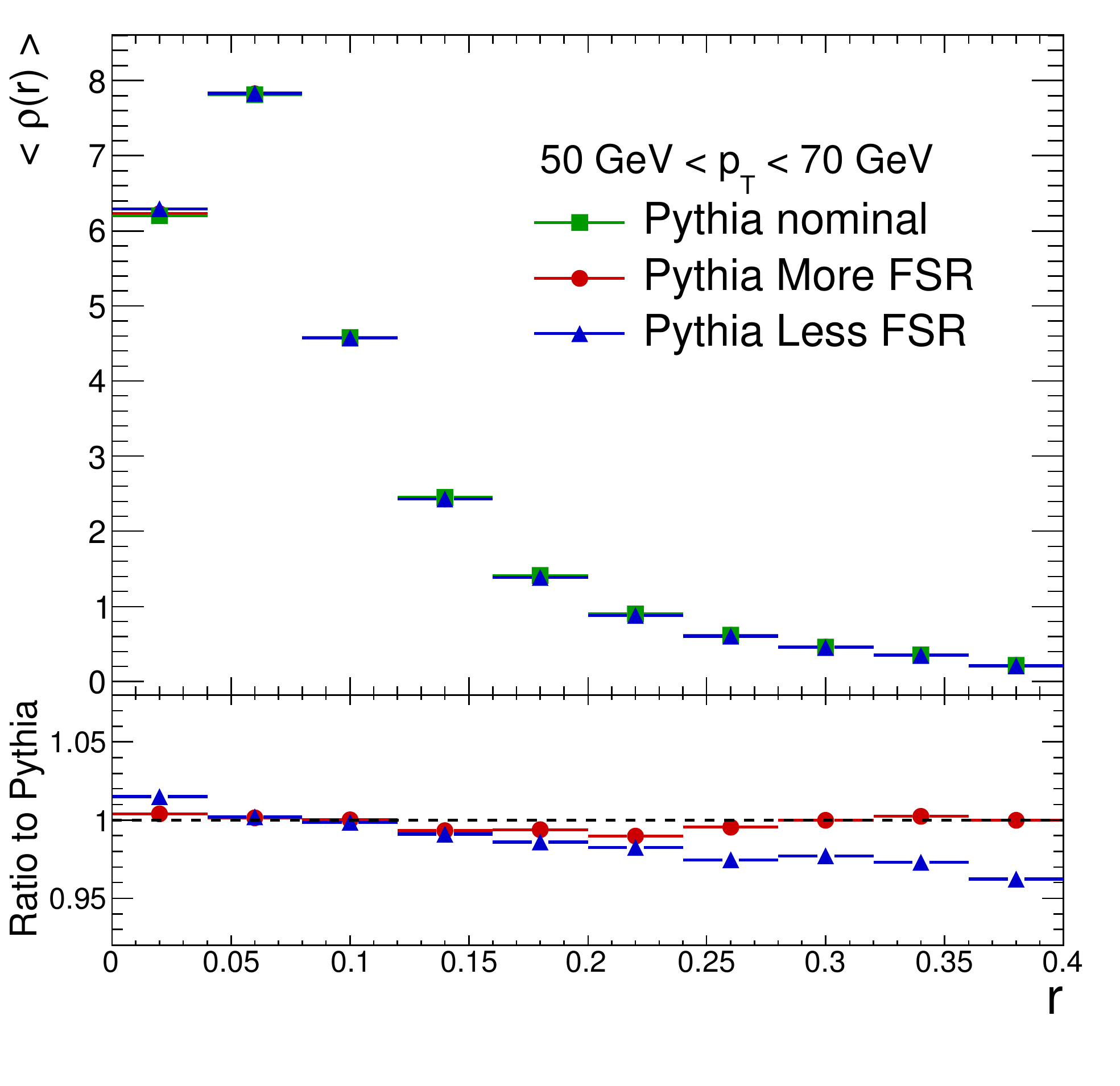}
\vspace{-1.cm}
\caption{The effects of the final-state radiation on the $b$-jet shapes.}
\label{fig:fsr}
\end{figure}

The effect of the amount of final-state radiation (FSR) on the $b$-jet shape distributions is studied by varying the parameters \textsc{PARP(72)} and \textsc{PARJ(82)}. These two parameters represent the value of $\Lambda_{\mathrm{QCD}}$ in the time-like showers responsible of the FSR (not arising from a resonant decay), and the infrared invariant mass cutoff, below which partons are not assumed to radiate. To increase the levels of FSR, these values are set to 0.384 and 0.5, respectively. To decrease the FSR activity, they are set to 0.096 and 2.0, respectively. This represents a change of a factor of 2 with respect to their nominal values 0.192 and 1.0. Fig. \ref{fig:fsr} shows the effect of these variations on the $b$-jet shapes, as well as the ratio to the nominal \textsc{Pythia} prediction. The impact of the FSR on the extracted $b$-quark mass is around $180 \MeV$.

\subsection{Colour reconnection}
\begin{figure}[H]
\centering
\includegraphics[width=5.5cm,height=5.9cm]{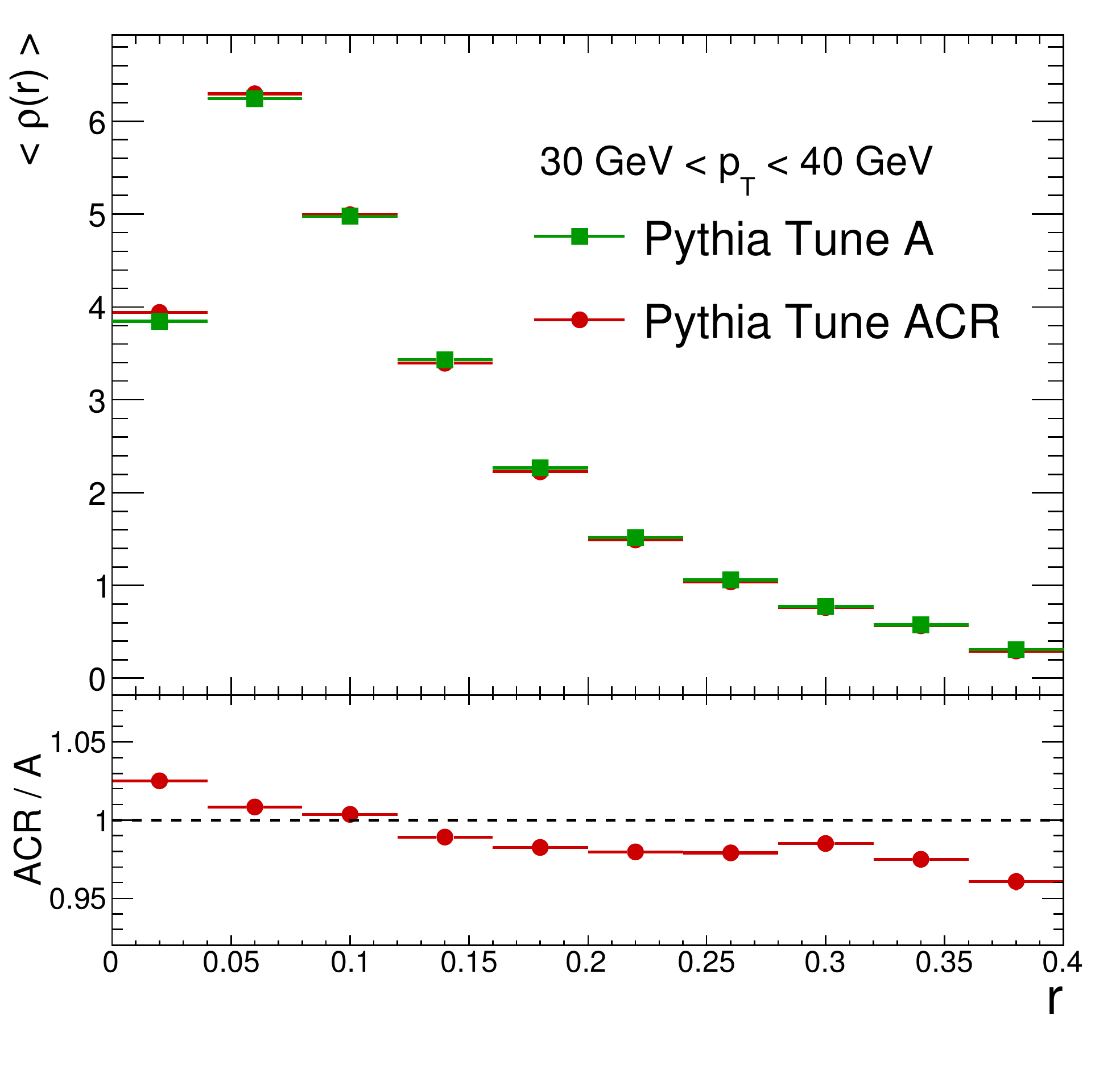}
\hspace{-0.25cm}
\includegraphics[width=5.5cm,height=5.9cm]{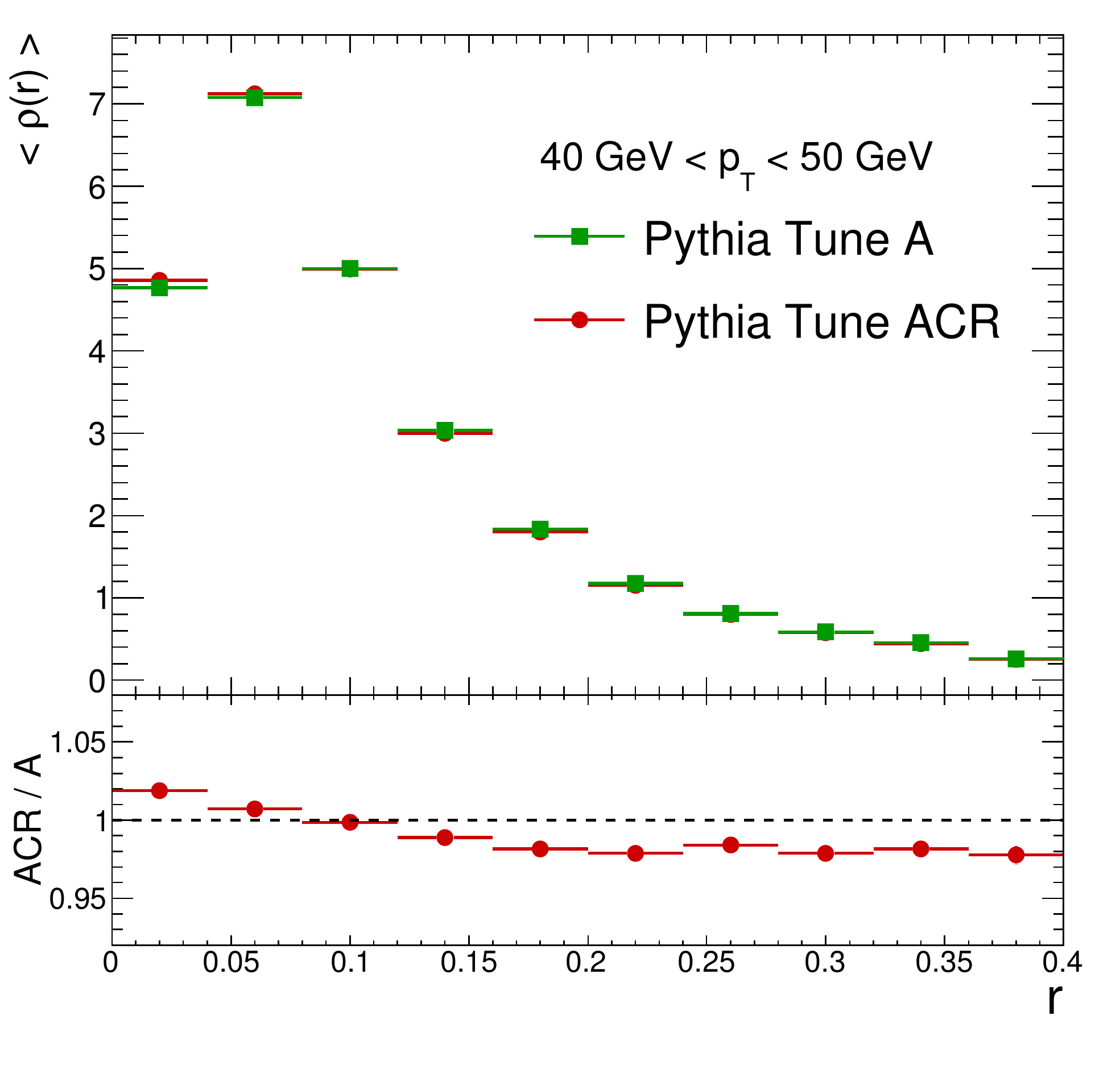}
\hspace{-0.25cm}
\includegraphics[width=5.5cm,height=5.9cm]{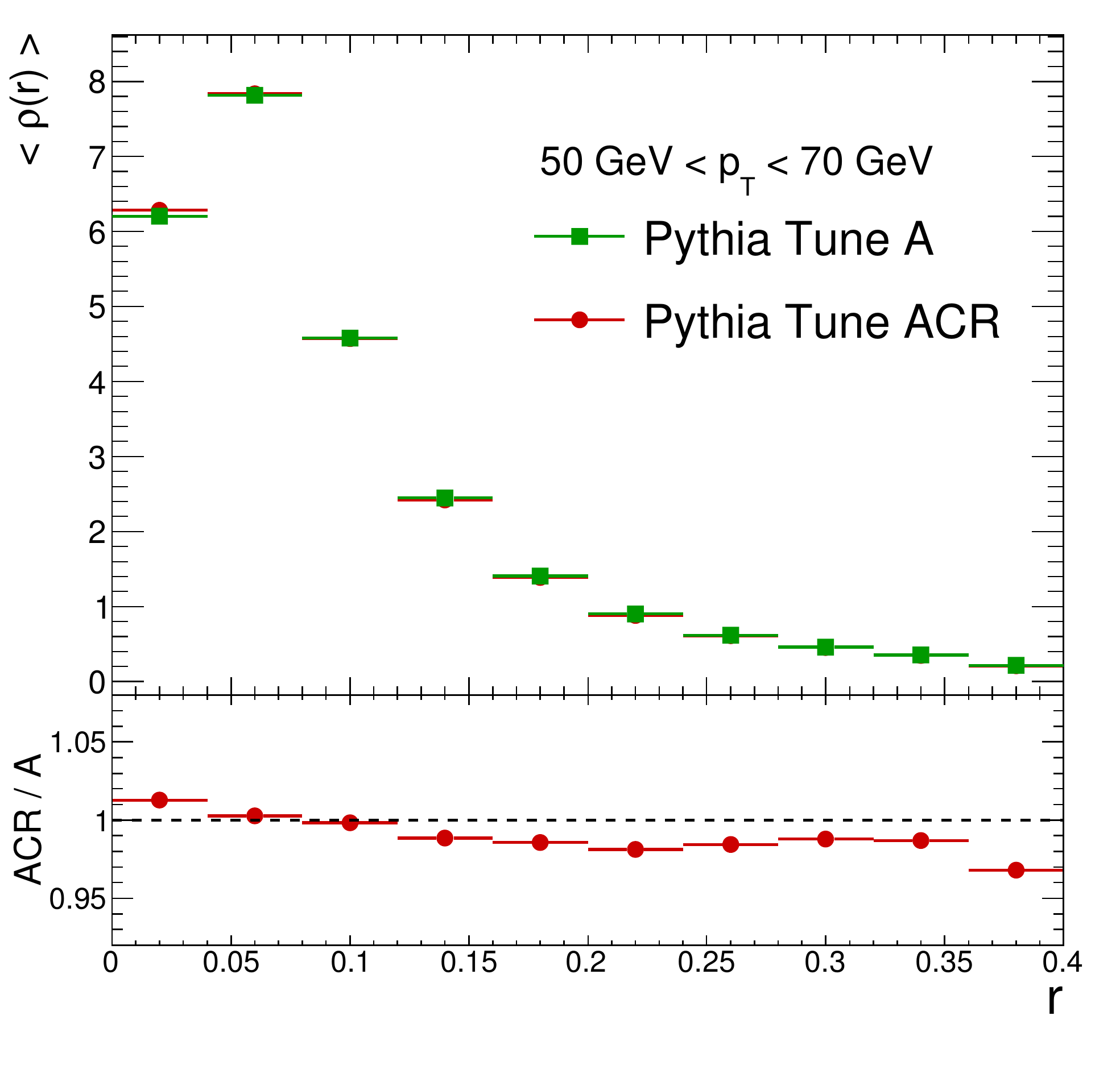}
\vspace{-1.cm}
\caption{The effects of the colour-reconnection modelling on the $b$-jet shapes.}
\label{fig:cr}
\end{figure}

The effect of the modelling of the colour reconnection (CR) between final-state partons is studied by using the \textsc{ACR} \cite{tuneACR} tuning of the \textsc{Pythia} Monte Carlo. This tune incorporates a new colour reconnection model, which assumes an enhanced amount of colour connections between partons with respect to the nominal \textsc{Tune A} sample. Figure \ref{fig:cr} shows that the effect of the new CR modelling is to increase the energy deposit on the jet cores on about 2\%. The impact on the $b$-quark mass is estimated by multiplying the nominal predictions by the ratio \textsc{Tune ACR}/\textsc{Tune A}, and it has an effect of around $170 \MeV$.

\subsection{Uncertainty on the $\Lambda_s$ determination.}
\label{secUncLambda}
The effects of the uncertainties in the determination of the parton shower scale have been also studied. To this end, the full set of $m_b$ variations have been generated again using the values of $\Lambda_s$ which define the envelope of its determination. Because the value obtained in section \ref{secLambda} was $\Lambda_s = 162.1 \pm 9.6 \MeV$, the full scan on $m_b$ variations has been repeated using the values $\Lambda_s = 152.5 \MeV$ and $\Lambda_s = 171.7 \MeV$, which define the endpoints of the interval in which $\Lambda_s$ can vary due to its experimental uncertainty. This is done in this way, instead of simply shifting each theoretical prediction by the nominal variation on the jet shapes due to this effect because the jet shapes are highly dependent on both parameters $m_b$ and $\Lambda_s$ at the same time. To keep track of this correlation, the full set of theoretical predictions has to be recalculated.\\
\newline
The fits with the varied values of $\Lambda_s$ are then repeated, and the differences between both of them and the central value are taken as the systematic uncertainties on the $b$-quark mass, which are in principle asymmetric. It is found that the impact on $m_b$ of the determination of the parton shower scale is around $60 \MeV$ at maximum, which represents the 1.2\% of the $b$-quark mass.\\
\newline
Another source of uncertainty related to the way in which $\Lambda_s$ is determined arises from the fixed order at which the running of $\alpha_s$ is evaluated. To estimate this uncertainty, the value of $m_b$ has been extracted using the running of $\alpha_s(Q^2)$ up to two loops, which is implemented for the \textsc{Herwig++} parton shower. The value of the two-loop shower scale was determined to be $\Lambda_s = 276.1 \pm 17.3 \MeV$, and the corresponding value of the $b$-quark mass is $m_b = 5.39 \pm 0.08 \GeV$. This value is to be compared with the value obtained for the one-loop running coupling in \textsc{Herwig++}, which was $m_b = 5.25 \pm 0.09 \GeV$, and therefore gives a relative uncertainty of 2.7\%. For the nominal value of $m_b = 4.86 \GeV$, this represents an additional uncertainty of $0.13 \GeV$, to be added in quadrature to the result of the propagation of the experimental uncertainty in $\Lambda_s$, and therefore has a maximum value of $0.14 \GeV$.\\
\newline
Discrepancies between data and MC on the description of the transverse momentum of light jets can lead to a biased result on the value of $\Lambda_{s}$. In order to check such effect, the light jet shapes were weighted and the fits were redone. These weights were estimated, in a very conservative way, matching the shape of the $\pt$ distributions of light and $b$-jets. The differences on $\Lambda_{s}$ with respect to the nominal value were found to be small ($\sim 3\%$). This difference is perfectly covered by the error on the fit procedure ($\sim 6\%$), thus ensuring the robustness of the $\Lambda_{s}$ determination.\\
\newline
As a further cross-check on the way in which $\Lambda_s$ is propagated throughout the analysis, the parton shower scale has been determined using the $b$-jet shapes obtained with the fitted value of $m_b$. The results are found to be fully compatible with the previous results obtained in Table \ref{tab:lambdaResults}, which reassures us on the extrapolation of $\Lambda_s$ from light-jets to $b$-jets.\\
\newline
After the evaluation of the theoretical uncertainties, the final value of the $b$-quark mass obtained in this analysis can be expressed as
\begin{equation}
m_b = 4.86 \pm 0.08 \mbox{ (exp.) } \pm 0.39 \mbox{ (Gen.) } ^{+0.02}_{-0.01} \mbox{ (ISR) }^{+0.18}_{-0.00} \mbox{ (FSR) } ^{+0.17}_{-0.00} \mbox{ (CR) } ^{+0.14}_{-0.13} \mbox{ (PS scale) }.
\label{eq:finalMb}
\end{equation}

\section{Summary and conclusions}
\label{secResults}
This study presents a determination of the mass of the $b$-quark using jet substructure techniques. It is found that the angular screening effects which were predicted in \cite{marchesini, dokshitzer} are confirmed and consistent with a reasonable value of the $b$-quark mass parameter. The dead cone effect was similarly exploited in \cite{perieanu} to determine the $c$-quark mass in $ep$ collisions at HERA.\\
\newline
Experimental uncertainties have been propagated using nuisance parameters for each source of uncertainty. This ensures that the correlations between all sources are explicitly taken into account. Systematic effects on the theoretical distributions have also been studied. The modelling of the jet shapes by different Monte Carlo generators is the main uncertainty on this analysis, accounting for an 8\% impact on the final value for $m_b$. Other systematic effects on the theoretical predictions have been studied, such as the amount of initial and final-state radiation, the colour reconnections and the uncertainty on the determination of the parton shower scale $\Lambda_s$. Our final result reads
\begin{equation}
m_b = 4.86 \pm 0.08 \mbox{ (exp.) } \pm 0.39 \mbox{ (Gen.) } ^{+0.02}_{-0.01} \mbox{ (ISR) }^{+0.18}_{-0.00} \mbox{ (FSR) } ^{+0.17}_{-0.00} \mbox{ (CR) } ^{+0.14}_{-0.13} \mbox{ (PS scale) }.
\end{equation}
Although there is a significant numerical similarity of this value with the value of the pole mass quoted by the Particle Data Group in \cite{pdg} and also with the values obtained by the LEP Collaborations in Refs. \cite{delphi} and \cite{aleph}, the value extracted here should not be confused with the QCD pole mass of the $b$-quark. It should rather be regarded as the on-shell mass parameter affecting the parton shower kinematics, as calculated in Ref. \cite{norr01}. In any case, it would be theoretically very interesting to define a way in which the Monte Carlo masses for hadronising quarks can be related to the poles of their respective fermionic propagators.

\paragraph{\textbf{Acknowledgements}}
The authors would like to thank Fernando Barreiro and Juan Terr\'on (UAM) for helpful discussions. We would also like to thank the ATLAS Collaboration for the many invaluable physics measurements already produced and the further to come. Communications with Torbj\"orn Sj\"ostrand (Lund) are also thankfully acknowledged.


\end{document}